\documentclass[twoside,twocolumn,9pt]{article}

\usepackage{authblk}
\usepackage{orcidlink}
\pdfoutput=1
\usepackage[utf8]{inputenc}
\usepackage[english]{babel}
\usepackage[T1]{fontenc}
\usepackage{amsmath}
\usepackage{amssymb}
\DeclareMathOperator{\Tr}{Tr}
\usepackage[separate-uncertainty=true]{siunitx}
\usepackage{makecell}
\usepackage{parskip}
\usepackage[left=1.5cm, right=1.5cm, top=1.785cm, bottom=2.0cm]{geometry}
\usepackage[format=plain,justification=justified,singlelinecheck=false,font={stretch=1.125,small,sf},labelfont=bf,labelsep=space]{caption}
\usepackage{sectsty}

\usepackage{tikz}
\usetikzlibrary{positioning, arrows.meta, shapes.geometric,fit}
\usepackage{lastpage}
\usepackage{lipsum}
\usepackage{listings}
\usepackage[numbers]{natbib}
\usepackage{xcolor}
\usepackage{booktabs}
\usepackage{float} 
\usepackage{fancyhdr}
\usepackage{fnpos}
\usepackage{algorithm}
\usepackage{algpseudocode}
\usepackage{hyperref}
\hypersetup{colorlinks=true, allcolors=blue}
\usepackage{setspace}
\usepackage[compact]{titlesec}

\usepackage[normalem]{ulem}
\usepackage[version=4]{mhchem}
\definecolor{cream}{RGB}{222,217,201}

\DeclareSIUnit{\angs}{\AA}
\DeclareSIUnit{\bohr}{a_0}
\DeclareUnicodeCharacter{2009}{\,}

\makeatletter
\newcommand{\abstractintitle}[1]{%
  \g@addto@macro\@maketitle{%
    \begin{center}
      \begin{minipage}{0.9\textwidth}
        \small #1 \vspace{3em}
      \end{minipage}
    \end{center}
    \vskip 1em}}
\makeatother

\begin{document}

\title{Molecular Dynamics with Nuclear Effects on Quantum Computers}

\author[1,2]{Lukas Haßfurth}
\author[1,2]{Juliane Heitkämper\orcidlink{0000-0003-2285-3381}}
\author[1,2]{Elias Walter\orcidlink{0000-0002-4206-0621}}
\author[1,2,3]{Birger Horstmann \orcidlink{0000-0002-1500-0578} \thanks{birger.horstmann@dlr.de}}

\affil[1]{\small German Aerospace Center (DLR), Wilhelm-Runge-Straße 10, Ulm, Germany}
\affil[2]{\small Helmholtz Institute Ulm, Helmholtzstraße 11, 89081, Ulm, Germany}
\affil[3]{\small Universität Ulm, Albert-Einstein-Allee 47, 89081, Ulm, Germany}

\date{}

\abstractintitle{
\normalsize
Nuclear quantum effects are critical for describing proton transfer and hydrogen bonding, but their incorporation into quantum chemistry calculations is often computationally prohibitive on classical hardware. 
A promising alternative are quantum computers due to their linear scaling in the space requirements with system size. We introduce a novel hybrid quantum-classical algorithm for ab-initio molecular dynamics that incorporates nuclear quantum effects via the nuclear-electronic orbital method. 
The proposed algorithm evaluates ground state energies and forces on the quantum computer using a variational quantum eigensolver, while the molecular geometries are updated classically. 
We validate our approach through simulations of \ce{H2}, \ce{H2O} and the Zundel ion \ce{H_5O_2^+}, comparing the simulated vibrational spectra with experimental data. Upon inclusion of nuclear quantum effects, the proton shuttling movement in the Zundel ion becomes effectively barrierless, and errors in the simulated frequencies improve significantly. 
Employing compact hardware-efficient ansatz circuits we achieve results comparable to the more accurate UCCSD ansatzes, which hints towards the feasibility of executing our algorithm on near-term quantum devices. 
}
\maketitle

\section{Introduction}

Nuclear quantum effects significantly influence many important biological and chemical processes.
For example, proton tunneling enhances reaction rates in proton transfer and proton-coupled electron transfer reactions such as photosynthesis \cite{Okamura2000, Berry2000}, respiration \cite{Saraste1999, Wikstrm1998,Michel1999} and general enzyme reactions \cite{Cha1989, VardiKilshtain2015}. 
They also influence the electrochemical nature of hydrogen bonds in water \cite{Tuckerman1997, Raugei2003, Kim2017} and ice \cite{Pamuk2012, Paesani2009}.

There exist various different frameworks for modelling quantum nuclei \cite{Nakai2007, Xu2020, Bochevarov2004, Cao1994}.
For many of these frameworks, incorporating nuclear quantum effects into quantum chemical calculations exponentially grows the full-configuration interaction (FCI) space, making exact calculations even for very small molecules infeasible on classical computers.

On quantum computers, the number of required qubits grows linearly with the system size, in principle making them a promising computational platform for the future, though significant quantum advantage is still to be demonstrated \cite{Weidman2024, Pfaendler2024}.

We therefore selected the nuclear electronic orbital (NEO) approach \cite{Webb2002} as one of the most promising candidates for implementation on a quantum computer.
In the NEO framework, selected nuclei are treated quantum-mechanically by placing them in nuclear orbitals, which are analogous to electronic atomic orbitals.
Then, coupled nuclear-electronic self-consistent field (SCF) equations are solved to obtain the set of molecular electronic and nuclear orbitals. 
This approach has the advantage that existing methods for purely electronic quantum chemistry on quantum computers can be extended naturally to include nuclear orbitals.

\begin{figure}
    \centering
    \includegraphics[width=\linewidth]{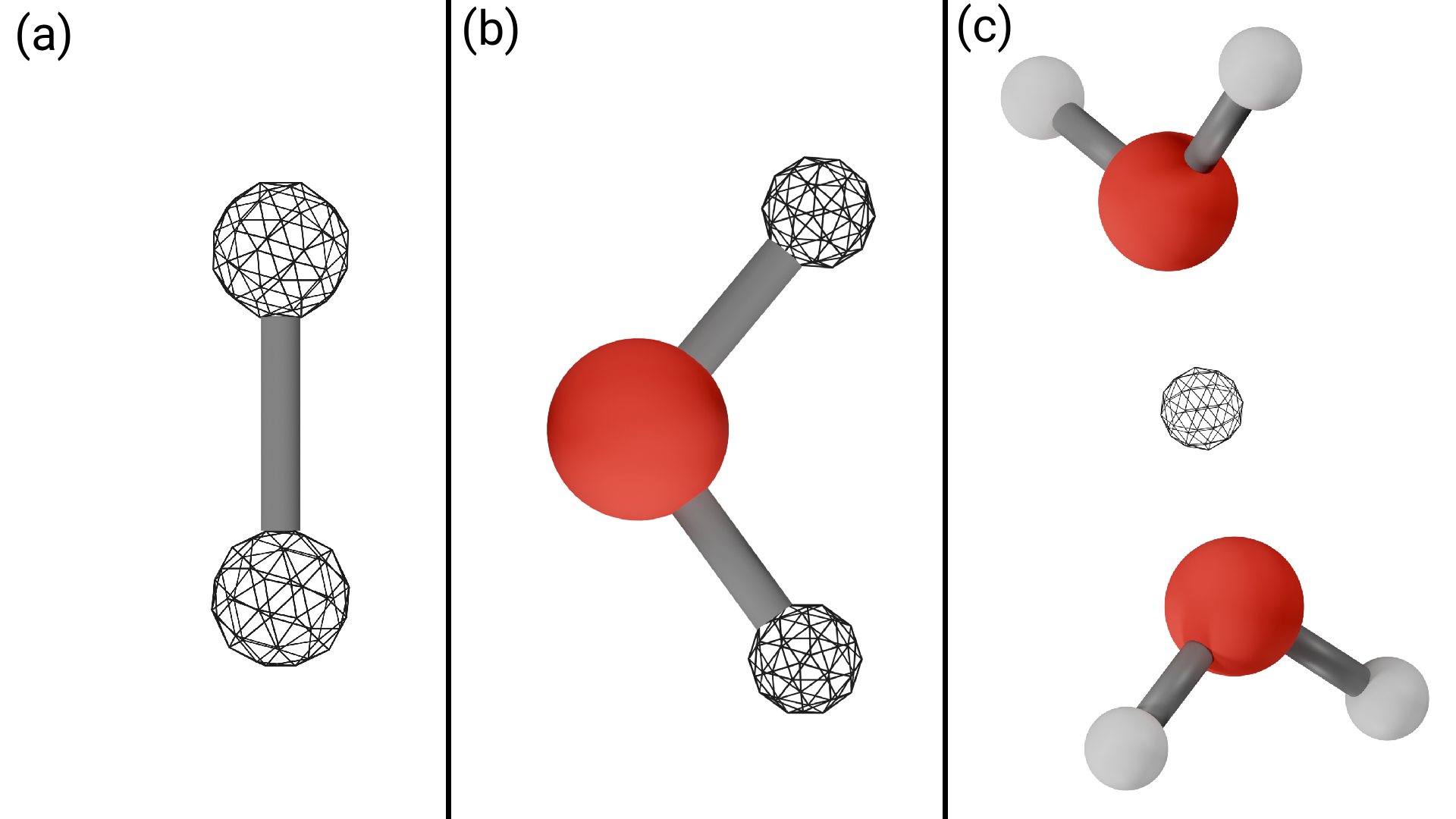}
    \caption{The investigated molecules. Nuclei that were treated with NEO are represented by the wireframes indicating isosurfaces of quantum nuclear probability density. (a) Hydrogen molecule. (b) Water. (c) Zundel ion.}
    \label{fig:overview_molecules}
\end{figure}

When studying more complex chemical processes, the simulation of molecular dynamics is often a method of choice \cite{Iftimie2005, Piccini2022, Yao2022}.
To demonstrate the feasibility of molecular dynamics (MD) simulations involving nuclear quantum effects on quantum computers, we implement a hybrid quantum-classical ab-initio molecular dynamics algorithm in this work. 
The algorithm uses the quantum computer as an efficient energy and force evaluator.
Our algorithm follows a second quantization approach and is based on the classical Born-Oppenheimer dynamics.
It evaluates the energy and the forces acting on the nuclei on the quantum computer and updates the molecular geometry on a classical computer, where selected nuclei are treated quantum-mechanically according to the NEO method.
After each geometry update, the new instantaneous ground state is found using the quantum computer.
While the search for the ground state can be implemented in various ways, we will employ the variational quantum eigensolver (VQE) algorithm \cite{Peruzzo2014} in this work.
We demonstrate the feasibility of the algorithm by simulating the vibrational spectra of the hydrogen molecule, the water molecule and the Zundel cation, which are depicted in Fig.~\ref{fig:overview_molecules}. The latter consists of two water molecules sharing a single proton, which serves as a minimal example for studying proton transfer \cite{Tuckerman1994, Agmon1995, Tuckerman1995}.

Similar algorithms have been proposed, but they either lack the inclusion of nuclear quantum effects \cite{Fedorov2021}, the simulation of dynamics \cite{Kovyrshin2023, Culpitt2025} or rely on pre-determined trajectories \cite{Kovyrshin2026}.
There also exist proposals for a first-quantization based MD simulation \cite{Kassal2008}, which may scale better for large systems.
However, while quantum computers still operate on the noisy intermediate scale era, they are restricted to simulating smaller systems where the second quantization approach is favorable.

This paper is organized as follows: In Sec.~\ref{sec:Methods}, we detail our methodological framework. 
In Sec.~\ref{sec:Results} we present and discuss our numerical results, comparing them with available experimental data for hydrogen, water, and the Zundel ion.
Finally, Sec.~\ref{sec:Summary} concludes this work by summarizing the key contributions.

\vspace{1em}

\section{Methods} \label{sec:Methods}

In Sec.~\ref{sec:Methods:BOMD} we introduce the molecular dynamics algorithm, which employs nuclear-electronic structure calculations to solve for the quantum-mechanical degrees of freedom (Sec.~\ref{sec:Methods:NEO}). 
We discuss how to reduce the number of quantum orbitals by means of active space techniques in Sec.~\ref{sec:Methods:Active_Spaces}, and how solve the quantum problem on a quantum computer in Sec.~\ref{sec:Methods:QC}.

\subsection{Quantum Born-Oppenheimer molecular dynamics} \label{sec:Methods:BOMD}
Our method for simulating ab initio molecular dynamics on a quantum computer is based on the work of Fedorov et al.~\cite{Fedorov2021}. The nuclei which are treated as classical point masses are propagated on the electronic potential energy surface (PES) following Hamiltonian mechanics:
\begin{align}
    \Dot{\mathbf{R}}_I &= \frac{\mathbf{P}_I}{M_I},\\
    \Dot{\mathbf{P}}_I = \mathbf{F}_I &= -\boldsymbol{\nabla}_I(E_0(R) + V_\text{NN}(R)),
\end{align}
where $\mathbf{R}_I$, $\mathbf{P}_I$, and $M_I$ refer to coordinate, momentum, and mass of the $I$-th nucleus. The force $\mathbf{F}_I$ is calculated as the negative gradient of the total energy with respect to the coordinates of the $I$-th nucleus. The total energy consists of the nuclear repulsion energy $V_\text{NN}(R)$ between the classical nuclei N and the ground state energy $E_0(R)$ of the electronic Hamiltonian, for given nuclear positions $R=\lbrace \mathbf{R}_I \rbrace_I$.

At every timestep, the instantaneous groundstate of the system needs to be prepared, which we obtain on a quantum computer using the variational quantum eigensolver (VQE) \cite{Peruzzo2014} (cf.~Sec.~\ref{sec:Methods:QC:VQE}). This guarantees that the Hellmann-Feynman theorem applies. The quantum force can be obtained by evaluation of the force operators on the ground state $\left| \psi_0(R) \right\rangle$: 
\begin{align}
    &\boldsymbol{\nabla}_I E_0(R) = \nonumber \\
    &=\sum_{k \in \{1,2,3\}} \mathbf{e}_k \frac{\partial}{\partial R_{Ik}} \langle \psi_0(R) | H(R) | \psi_0(R)\rangle \nonumber 
    \\ &=\sum_{k \in \{1,2,3\}} \mathbf{e}_k \left\langle \psi_0(R) \left| \frac{\partial H(R)}{\partial R_{Ik}} \right| \psi_0(R) \right\rangle\nonumber \\
    &=  \left\langle \psi_0(R) \left| \boldsymbol{\nabla}_I  H(R) \right| \psi_0(R) \right\rangle,
\end{align}

where $\mathbf{e}_k$ is the $k$-th unit vector and $H(R)$ the electronic or NEO Hamiltonian.

In summary, a timestep at the discretized time $t_i$ in the proposed ab-initio molecular dynamics (AIMD) algorithm consists of the following steps, where we use the notation $f(t_i) := f(R(t_i))$:

\begin{enumerate}
    \item Given $R(t_i)$, obtain the Hamiltonian $H(t_i)$ and the force operators $\boldsymbol{\nabla}_I H(t_i)$ for each nuclear coordinate.
    \item Find the groundstate $|\psi_0(t_i)\rangle$ and the groundstate energy $E(t_i)$ with VQE, i.e.\ by finding the parameters $\theta$ of some suitable ansatz $U(\theta)|0\rangle$ for which $E(t_i)$ is minimal.
    \item Measure the expectation values of the force operators and add the nuclear repulsion force to acquire the total force on each nuclear coordinate, i.e. 
    \begin{align}
        \mathbf{F}_{I}(t_i) &= -\left\langle \psi_0(t_i) \left| \boldsymbol{\nabla}_I H(t_i) \right| \psi_0(t_i)\right\rangle \nonumber\\ 
        &-\boldsymbol{\nabla}_I V_\text{NN}(R)
    \end{align}
    \item Update the nuclear coordinates. We use the velocity Verlet algorithm \cite{Swope1982}:
    \begin{align}
        \mathbf{R}_I(t_{i+1}) &= \mathbf{R}_I(t_i) + \mathbf{V}_I(t_i) \Delta t + \frac{\mathbf{F}_I(t_i)}{2M_I} \Delta t^2 \\
        \mathbf{V}_I(t_{i+1}) &= \mathbf{V}_I(t_i) + \frac{\mathbf{F}_I(t_i) + \mathbf{F}_I(t_{i+1})}{2M_I} \Delta t
    \end{align}
\end{enumerate}

\tikzset{
    start/.style={
        ellipse,
        rounded corners,
        draw=black,
        fill=red!10,
        line width=1pt,
    },
    data/.style={
        trapezium,
        trapezium left angle=70,
        trapezium right angle=-70,
        rounded corners,
        draw=black,
        fill=blue!10,
        line width=1pt,
    },
    process/.style={
        rectangle,
        draw=black,
        fill=yellow!10,
        line width=1pt,
    },
    decision/.style={
        diamond,
        draw=black,
        fill=orange!10,
        line width=1pt,
    },
    arrow/.style={
        -{Latex[length=3mm]},
        thick
    }
}
\begin{figure*}
\centering

\begin{tikzpicture}
\useasboundingbox (0,0) rectangle (\textwidth, 9.5cm);
\node[start] (start) at (\textwidth/2- 4.5cm,9cm) {Start};
\node[data,right=of start] (input_data) {Current molecular geometry};
\node[process, below=of input_data] (vqe) {Solve VQE};
\node[data,below=of vqe] (vqe_result) {Energy and forces};
\node[decision,below=of vqe_result] (final_check) {Final time reached?};
\node[process, right=of vqe] (integration) {Propagate atomic positions};
\node[data,left=of final_check](trajectory_data) {Trajectory};
\node[start,above=of trajectory_data](end) {End};

\draw[arrow] (start) -- (input_data);
\draw[arrow] (input_data) -- (vqe);
\draw[arrow] (vqe) -- (vqe_result);
\draw[arrow] (vqe_result) -- (final_check);
\draw[arrow] (final_check) -- node[pos=0.4,above] {Yes} (trajectory_data);
\draw[arrow] (final_check.east) -|  node[pos=0.1,above] {No} (integration.south);
\draw[arrow] (integration) |- (input_data);
\draw[arrow] (trajectory_data) -- (end);

\node[draw=blue, dashed, fit=(vqe) (vqe_result),label={[blue,label distance=0cm,align=left]180:Quantum\\Computer}] {};
\end{tikzpicture}
\caption{Schematic representation of the hybrid-classical MD algorithm.}
\label{fig:md_algorithm}
\end{figure*}

A concise version of the algorithm is given in Algorithm \ref{alg:VQE_MD}. Fig.~\ref{fig:md_algorithm} shows the algorithm as a flowchart.
\begin{algorithm}
\caption{VQE-based molecular dynamics}\label{alg:VQE_MD}
\begin{algorithmic}[1]
\Require $R(t_0)$, $U(\theta)$
\State $V(t_0) \gets 0$

\While{$t_i < t_{\text{end}}$}

\State $\theta(t_i) \gets \arg \min_{\theta} \langle 0|  U^\dagger(\theta)  H(t_i) U(\theta) |0\rangle$

\State $|\psi_0(t_i)\rangle \gets U(\theta(t_i))|0\rangle$

\State $F_{I k}(t_i) \gets  -\left\langle \psi_0(t_i) \left| \frac{\partial H(t_i)}{\partial R_{I k}} \right| \psi_0(t_i)\right\rangle$ \newline \hspace*{6em} $- \frac{\partial}{\partial R_{I k}} V_{\text{NN}}$

\State $\mathbf{R}_I(t_{i+1}) \gets \mathbf{R}_I(t_i) + \mathbf{V}_I(t_i)\Delta t + \frac{\mathbf{F}_I(t_i)}{2M_I} \Delta t^2$

\State $\mathbf{V}_I(t_{i + 1}) \gets \mathbf{V}_I(t_i) + \frac{\mathbf{F}_I(t_i) + \mathbf{F}_I(t_{i+1})}{2 M_I}\Delta t$
\EndWhile
\end{algorithmic}
\end{algorithm}

\subsection{Nuclear-electronic structure calculations} \label{sec:Methods:NEO}
\subsubsection{Energy estimation}
The nuclear-electronic orbital (NEO) theoretical framework \cite{Webb2002} does not apply the standard Born-Oppenheimer separation between electrons and all nuclei and the oftentimes subsequent classical treatment of the nuclei.
Instead, selected light nuclei can be treated in second quantization and on the same footing as the electrons.
While for this work we assume that all quantum nuclei are protons, we note that the NEO framework does not limit which nuclei can be treated quantum mechanically.
The NEO approach then leads to coupled nuclear-electronic Hartree-Fock (NEO-HF) equations. 
Solving the NEO-HF equations yields molecular orbital (MO) coefficients $\{C^\text{e}, C^\text{n}\}$ for electronic and nuclear orbitals $\phi_p(\mathbf{r})$ and $\phi_P(\mathbf{r})$, respectively. These transform the atomic electronic and nuclear orbitals $\chi_\mu^\text{e}$ and $\chi_\mu^\text{n}$ into MOs
\begin{align}
    \phi_p(\mathbf{r}) &= \sum_\mu C^\text{e}_{\mu p} \chi^\text{e}_\mu (\mathbf{r}) \\
    &\text{and}\nonumber\\
    \phi_P(\mathbf{r}) &= \sum_\mu C^\text{n}_{\mu P} \chi^\text{n}_\mu (\mathbf{r}),
\end{align}
where electronic MOs are indexed by $\{p,q,r,s\}$, while the capitalized indices $\{P,Q,R,S\}$ refer to the nuclear MOs.

The NEO Hamiltonian in second quantization is then
\begin{align}
\label{eq:hamiltonian}
    H_{\text{NEO}} &= \sum_{pq} h^\text{e}_{pq} a^\dagger_p a_q + \frac{1}{2} \sum_{pqrs} g^\text{e}_{pqrs} a^\dagger_p a^\dagger_r a_s a_q \nonumber \\
    &+\sum_{P Q} h^\text{n}_{P Q} a^\dagger_{P} a_{Q}  \nonumber \\ 
    &+ \frac{1}{2} \sum_{P Q R S } g^\text{n}_{P Q R S} a^\dagger_{P} a^\dagger_{R} a_{S} a_{Q} \nonumber \\
    &- \sum_{pq}\sum_{P Q} g^\text{en}_{pqPQ} a^\dagger_{p} a^\dagger_{P} a_Q a_{q}.
\end{align}

The one-body electronic and nuclear integrals are
\begin{align}
    \label{eq:hcore_el_definition}
    h^\text{e,n}_{pq} &= \int \phi_p^*(\mathbf{r})\Bigg[-\frac{1}{2 m^\text{e,n}}\nabla^2 \nonumber \\
    &- e^\text{e,n} \sum_I \frac{Z_I}{\left| \mathbf{R}_I - \mathbf{r}\right|}\Bigg]\phi_q(\mathbf{r})d^3r,
\end{align}

where $I$ iterates over all classically-treated nuclei, while $Z_I$ and $\mathbf{R}_I$ are the atomic numbers and positions of these nuclei, respectively. 
The superscripts $\text{e}$ and $\text{n}$ indicate electronic and nuclear quantities, respectively. 
The charge and mass of electrons and protons are given in atomic units as $e^\text{e}=1$, $e^\text{n}=-1$, $m^\text{e}=1$ and $m^\text{n}=1836$.

The two body integrals are
\begin{equation}
        \label{eq:eri_el_definition}
        g^\text{e}_{pqrs} = \int \frac{\phi_p^*(\mathbf{r}_1)\phi_q(\mathbf{r}_1)\phi^*_r(\mathbf{r}_2)\phi_s(\mathbf{r}_2)}{\left|\mathbf{r}_1 - \mathbf{r}_2\right|}d^3r_1 d^3r_2, 
\end{equation}
and $g^\text{p}_{PQRS}$ and $g^\text{en}_{pqPQ}$ defined equivalently by replacing 2 or 4 lower-case indices by upper-case indices, respectively.

\subsubsection{Evaluation of energy gradients}
As discussed previously, the nuclear gradient of the Hamiltonian is required to evaluate the forces necessary for molecular dynamics.
An analytical expression for $dE_\text{NEO-HF}/dR$ in the MO basis involving AO integral derivatives has already been derived in Ref.~\cite{Webb2002}. By splitting this expression into an operator and the density with which it is contracted, a formulation for the nuclear gradient of the NEO-Hamiltonian $dH_\text{NEO}/dR$ can be obtained. We list this operator in Appendix~\ref{sec:gradient_operators} for reference.

Alternatively, the gradient can be derived using electronic structure theory derivative theory techniques and their analogous expansion to NEO. In Ref.~\cite{OBrien2022}, this is done to directly obtain the nuclear gradient of the purely electronic Hamiltonian.

While these gradient operators are derived for the NEO and purely electronic case, respectively, we can still compare their general structure. Their main structural difference is that the derivative in Ref.~\cite{Webb2002} exploits the symmetries of the one- and two-body integrals $h_{pq}$, $g_{pqrs}$ which Ref.~\cite{OBrien2022} does not, enabling a more compact formulation of the former.
These two formulations of the gradient operators are only equivalent under certain conditions. 
We will elaborate this further when we introduce the active space transformation in Sec.~\ref{sec:active_space_transformation}, and show detailed calculations in App.~\ref{sec:compare_force_ops}.

\subsubsection{NEO dynamics}
We adopt the semi-classical approximation from Ref.~\cite{Zhao2020} to treat the nuclear basis centers as the mean positions of each quantum nucleus, with the consequence that the nuclear basis set centers move according to the forces acting on the quantum nuclei. 
By including this approximation, Alg.~\ref{alg:VQE_MD} remains valid for both purely electronic and NEO treatment of the molecular systems.
This requires a one-to-one correspondence between quantum nuclei and basis set centers, which does not allow for the use of auxiliary basis sets.
This approximation is only valid as long as the quantum nuclei remain localized and quasi-distinguishable particles. 
As we demonstrate in Fig.~\ref{fig:basisset_dependence_h2} of Appendix~\ref{sec:app_basis_set_dependence_neo}, simulations with larger nuclear basis sets severely underestimate the simulated frequencies.

This is because larger nuclear basis sets allow quantum nuclei to approach their equilibrium positions without actually moving the nuclear basis set centers.
Conversely, this also means that the nuclear density is less dependent on the concrete positions of the nuclear basis set centers.
As the quantum parts of the forces on the nuclei are calculated from the energy response of basis set repositioning, this lower sensitivity causes an underestimation of the acting forces.

Concretely, the additional orbitals of basis sets with higher angular momenta allow the quantum nuclei to move relative to the basis set centers. 
While this additional flexibility is usually beneficial, our algorithm requires the quantum nuclei to be pinned to the nuclear basis set centers. 
We share this restriction with the CNEO\cite{Xu2020}-based approach, but instead of 
additional Lagrange multipliers in the NEO-SCF equations and penalty terms in the energy functional, we implement it by excluding nuclear basis functions with angular momentum.
Therefore, the coincidence of basis set centers and nuclear expectation values needs to be checked for every system under study.

\subsection{Active Spaces} \label{sec:Methods:Active_Spaces}
\subsubsection{Active Space Transformation}
\label{sec:active_space_transformation}

The limited number of qubits available on noisy intermediate-scale quantum computers (NISQ) restricts the maximum problem size that can be treated. 
We reduce the problem size on the quantum computer by implementing the Hartree-Fock embedding described in Ref.~\cite{Rossmannek2021}, where an active space is selected which is treated on the quantum computer while the remaining orbitals are treated at the Hartree-Fock level. 

In the following, $\{i,j\}$ denote the indices of inactive electronic MOs which are occupied according to the HF solution, $\{p,q,r,s\}$ denote general electronic MOs and $\{u,v,x,y\}$ denote active electronic MOs. The capitalized versions of these indices refer to their nuclear counterparts.

Generalizing the embedding to NEO yields the active space NEO Hamiltonian $H_{\text{NEO}}^\text{A}$:
\begin{align}
    H_{\text{NEO}}^\text{A} &= E^\text{I} + \sum_{uv} F^\text{I(e)}_{uv} a^\dagger_u a_v +\sum_{U V} F^\text{I(n)}_{U V} a^\dagger_{U} a_{V} \nonumber \\
    &+\frac{1}{2} \sum_{uvxy} g^\text{e}_{uvxy} a^\dagger_u a^\dagger_x a_y a_v \nonumber \\
    & + \frac{1}{2} \sum_{U V X Y } g^\text{n}_{U V X Y} a^\dagger_{U} a^\dagger_{X} a_{Y} a_{V} \nonumber \\
    &- \sum_{uv}\sum_{U V} g^\text{en}_{u v U V} a^\dagger_{u} a^\dagger_{U} a_V a_{v}.
    \label{eq:embedding_Hamiltonian}
\end{align}

The inactive energy $E^\text{I}$ of the always occupied electronic and nuclear orbitals is given by
\begin{align}
    E^\text{I} &= \sum_{j} \left( h^\text{e}_{jj} + F^\text{I(e)}_{jj} \right) + \frac{1}{2} \sum_{J} \left( h^\text{n}_{J J} + F^\text{I(n)}_{J J} \right).
\end{align}
The inactive Fock operators for the electronic and nuclear subspace $F^\text{I(e)}_{pq}$ and $F^\text{I(n)}_{P Q}$, respectively, are conventionally defined as
\begin{align}
    F^\text{I(e)}_{pq} &= h^\text{e}_{pq} + \sum_{i} (2  g^\text{e}_{iipq} - g^\text{e}_{iqpi}) - \sum_{I} g^\text{en}_{pq I I} \nonumber \\
    F^\text{I(n)}_{P Q} &= h^\text{n}_{P Q} + \sum_{I} (g^\text{n}_{I I P Q} - g^\text{n}_{I Q P I}) - 2 \sum_{i} g^\text{en}_{iiP Q} .
    \label{eq:inactive_Fock_short}
\end{align}

Here, we use the more general form of the inactive Fock operators
\begin{align}
    F^\text{I(e)}_{pq} &= h^\text{e}_{pq} - \sum_{I} g^\text{en}_{pqI I} \nonumber \\
    &+ \frac{1}{2}\sum_i \left(2g^\text{e}_{iipq} - g^\text{e}_{iqpi} + 2g^\text{e}_{pqii} - g^\text{e}_{qiip}\right) \nonumber\\ 
    F^\text{I(n)}_{P Q} &= h^\text{n}_{P Q} - 2\sum_{i} g^\text{en}_{iiP Q} \nonumber \\
    &+ \frac{1}{2}\sum_i \big(g^\text{n}_{I I P Q } - g^\text{n}_{I Q P I} + g^\text{n}_{P Q I I} - g^\text{n}_{Q I I P}\big) .
    \label{eq:inactive_Fock_long}
\end{align}

While Eqs.~\eqref{eq:inactive_Fock_short} and \eqref{eq:inactive_Fock_long} are equivalent for the calculation of energies, it is beneficial to write the inactive Fock operator in the more general form with regard to the gradient calculation. 
As we show in App.~\ref{sec:gradient_operators} and  \ref{sec:compare_force_ops}, differentiating the more general Fock operator directly generates the terms to correctly include the MO response to the nuclear motion of both active and inactive orbitals.

\subsubsection{Active Space Selection}
\label{sec:ActiveSpaceSelection}
A good selection of active space orbitals is crucial for accurate calculations. 
The popular autoCAS \cite{autocas} active space finding method provides a principled and automatic approach to finding active spaces, but produces spaces that are very large for current NISQ devices as both the electrons and selected nuclei need to be mapped to qubits.
Other approaches, like the active space finder by HQS \cite{ActiveSpaceFinder}, work with natural orbitals. As our gradient expressions are derived for canonical orbitals, they are currently incompatible with this latter set of methods.
To address these issues, we combine three criteria with the goal of finding a minimum viable active space.

We score orbitals first according to their single-orbital entropies \cite{Stein2017} and then by their variability along the expected reaction path. 
The heuristic justification for the variability criterion is that orbitals which participate strongly in a reaction are also well suited for its description.

For the calculation of the variability score, the process trajectory is divided into $N$ images. Then, the score $r_k$ of each orbital $k$ is the accumulated difference between unity and the overlap $S_{k,k}$ of each orbital with its corresponding orbital in the previous image:
$$
r_k = \sum_{i=1}^N \left( 1 - S_{k,k}^{(i-1,i)} \right).
$$

The third criterion enforces symmetry and is ensuring that the active space is invariant under the chemical reaction’s geometric transformations. For instance, in proton transfer reactions, the system often exhibits reflection symmetry across the reaction coordinate. 
To preserve this physical symmetry, we require that if an orbital is included in the active space, its mirror image (with respect to the symmetry plane) must also be included. 
This ensures that the active space remains physically consistent throughout the reaction and avoids unphysical asymmetries.

Based on these three criteria including the discussed symmetry considerations, we construct chemically well-informed, compact and high-quality active spaces.

\subsubsection{Corresponding Orbitals}

Defining the active space solely by the energetic indices of the MOs can cause MOs to swap in and out of the active space during dynamic processes, as the changing geometry may cause energetic reordering. Instead, we additionally define a reference geometry to uniquely fix the active space. 

The MO indices of the active space for a different geometry can then be obtained by maximizing the overlap integrals with the active space on the reference geometry $R_{\text{ref}}$.
Let the intermolecular overlap $\tilde{S}$ be
$$
\tilde{S}_{il} = \sum_{jk} C^{\text{ref}}_{ij} \langle \chi_j(r; R_{\text{ref}}) | \chi(r; R_{\text{new}}) \rangle  C^{\text{new}}_{kl}.
$$

Then, $\tilde{S}_{il}$ is the overlap between MO $i$ on the reference geometry with MO $j$ on the new geometry.
Finding the globally optimal mapping between these two sets of MOs is then achieved by solving the linear assignment problem employing $-\tilde{S}$ as the cost function.
Once the mapping is established, it can be used to identify the active space on the new molecular geometry $R_{\text{new}}$.

\subsection{Quantum Chemistry on Quantum Computers} \label{sec:Methods:QC}

\subsubsection{Fermion-to-Qubit mappings}

The fermionic representation of the Hamiltonian (\ref{eq:hamiltonian}) needs to be mapped to the qubits of the quantum computer. 
We use the parity mapping scheme \cite{Seeley2012}, since it allows the trivial removal of two redundant qubits due to particle number conservation.

If the parity mapping is extended to NEO \cite{Kovyrshin2023}, two additional qubits can be tapered, since the number of quantum nuclei is also preserved.
Because we assume a spin-maximized nuclear state, the nuclear beta-spin sector can be projected out before the fermion-to-qubit mapping. This causes a further reduction in the qubit count requirements.

In principle, it is possible to employ Z2 tapering \cite{Bravyi2017} to exploit the point group symmetries of the molecular system.
During molecular dynamics, these symmetries are subject to change, potentially causing an inconsistent number of qubits between timesteps.
We therefore leave the additional Z2 tapering for a future work.

\subsubsection{Variational Quantum Eigensolver} \label{sec:Methods:QC:VQE}

Our proposed MD algorithm relies on the assumption that the instantaneous groundstate of the molecular system can be prepared on the quantum computer. 
In general, the state of the quantum computer can be manipulated by applying a sequence of gates, which may also be called a quantum circuit $U$. If the sequence of applied gates is fixed, the remaining freedom of the quantum circuit is the choice of gate parameters $\{\theta_i\}_{i=0}^n$. This combination of a fixed gate structure with variable gate parameters is called an ansatz $U(\boldsymbol{\theta})$.

The variational quantum eigensolver (VQE) is an algorithm that minimizes  the expectation value of a Hamiltonian with respect to the parameters of a given ansatz. The expectation value for a given set of parameters is used as the score for a classical optimizer, thereby requiring evaluation of the expectation value on the quantum computer for each optimization step. The optimized parameters can then be used to recreate the ground state and evaluate the expectation values of other observables, that is,
\begin{align}
    E^* &= \min_{\boldsymbol{\theta}} \langle 0 | U^\dagger(\boldsymbol{\theta}) H U(\boldsymbol{\theta}) | 0 \rangle \\
    \boldsymbol{\theta}^* &= \arg\min_{\boldsymbol{\theta}} \langle 0 | U^\dagger(\boldsymbol{\theta}) H U(\boldsymbol{\theta} | 0 \rangle \\
    \langle A \rangle_{\boldsymbol{\theta}^*} &= \langle 0 | U^\dagger(\boldsymbol{\theta}^*) A U(\boldsymbol{\theta}^*) | 0 \rangle 
\end{align}
for an arbitrary observable $A$.

The choice of ansatz greatly influences the efficacy of the VQE algorithm. 
Thus, we close this section with a short introduction of relevant ansatz types.

The unitary coupled cluster ansatz with singles and doubles (UCCSD) \cite{Barkoutsos2018} is chemically motivated. 
While providing generally good results for chemical systems, its circuit is typically too long to be feasible on NISQ devices. 

There is another class of ansatzes that are focused on being applicable on current quantum devices while still being able to approximate the target state. These are called hardware-efficient ansatzes. 
One such ansatz is the tiled unitary product ansatz (tUPS) \cite{Burton2024}. 

Another hardware-efficient ansatz is the real amplitudes ansatz \cite{Kandala2017}, which consists of alternating layers of Ry rotations and CNOT entangling gates.
The gates are chosen in such a way that the final quantum state is purely real.

Both of these hardware-efficient ansatzes are designed to be repeated multiple times to create a layered structure. 
A larger number of layers increases the expressivity but also the execution cost of the circuit.

\subsection{Computational Details}

All our calculations employ the electronic 6-31G basis set \cite{Hehre1972}.
Additionally, we define two nuclear basis sets which are derived from the DZSPDN basis set \cite{Webb2002}.
The first set, denoted as the double-zeta $s$ nuclear basis set (DZSNB), is obtained by discarding all shells from the DZSPDN basis set that are not $s$-type.
The second set, which we call the double-zeta $s,p$ nuclear basis set (DZSPNB), is constructed similarly by retaining only the $s$- and $p$-type shells.
For both of the derived nuclear basis sets, the coefficients were not changed relative to the DZSPDN basis set.

In the following, we investigate molecular hydrogen \ce{H2}, water \ce{H2O}, and the Zundel cation \ce{H5O2+}.
For NEO calculations, certain nuclei are treated as quantum particles. 
As can be seen in Fig.~\ref{fig:overview_molecules}, all hydrogen nuclei are selected for \ce{H2} and \ce{H2O}, while for \ce{H5O2+}, only the central hydrogen is selected for quantum treatment.
The DZSNB basis set was used for all NEO calculations for \ce{H2} and \ce{H2O}. 
For the Zundel ion, the effect of the nuclear basis sets on the simulated dynamics was assessed by performing separate NEO calculations with the DZSNB and the DZSPNB basis set.

While it is possible to simulate the \ce{H2} trajectory with all orbitals mapped onto the quantum computer within reasonable time, the other systems require an active space reduction.
We determine an electronic active space for water consisting of 5 orbitals and 6 electrons, while for the Zundel ion we select an active space of 5 orbitals and 4 electrons.
The details of the active spaces are described in their respective chapters.

We utilize the parity transformation with particle number tapering for mapping between the fermionic and the qubit space in all simulations.

Unless otherwise stated, the variational circuit is built from the UCCSD ansatz for purely electronic problems and the NEOUCCSD \cite{Kovyrshin2023} ansatz for NEO problems.
We chose the COBYLA algorithm \cite{Zhang_2023} for classical optimization, based on a small set of test optimizations comparing different optimizers.

For every system, we simulate a trajectory with a length of \SI{1}{\pico \second} and a fixed timestep of \SI{121}{\atto \second}, which corresponds to a final frequency resolution of \SI{33}{\per \centi \metre}.

We employ a custom in-house calculation protocol written in the Python programming language based on the Atomic Simulation Environment \cite{HjorthLarsen2017} \texttt{Calculator} class for the preprocessing of the molecular systems. 
This calculator implements the hybrid quantum-classical energy and force evaluation described above, where the noiseless statevector simulator provided by Qiskit \cite{qiskit2024} version \texttt{1.3.2} facilitates preparation and measurement of states on a simulated quantum computer.
The purely electronic and NEO SCF procedures are written from scratch, employing PySCF \cite{Sun2017} version \texttt{2.7.0} as the integral generator and NumPy \cite{harris2020array} version \texttt{1.26.4} for accelerating linear algebra operations.
The initial ground state geometries are obtained by using our calculator in conjunction with the FIRE2 geometry optimizer \cite{Gunol2020} on the classical (NEO)FCI or (NEO)CASCI potential energy surfaces.
The active spaces for this optimization are the same as presented below for the dynamical simulations.
These optimized geometries can be found in Appendix \ref{sec:app_equilibrium_geometries}.

After determination of the groundstate geometries, we calculate the Hessians for each system and perform a normal mode analysis.
From these normal modes, we select
\begin{itemize}
    \item for \ce{H2}: the \ce{H-H} stretch mode,
    \item for \ce{H2O}: the bend, symmetric stretch and antisymmetric stretch mode (see Fig.~\ref{fig:normal_modes_water}) and
    \item for \ce{H5O2+}: the proton transfer mode
\end{itemize}
as test cases for our molecular dynamics algorithm.
For every selected normal mode, we run a simulation where the optimized geometry of the molecule was perturbed along that mode by an energy of \SI{50}{\milli \eV}. 
For these simulations, the calculator uses the hybrid quantum-classical approach for calculating energies and forces.

As our simulation takes into account anharmonic effects due to working on ab-initio potential energy surfaces, it is not obvious which initial energy best matches experimental conditions.
Therefore, the initial excitation energy is chosen on the basis of a quantum wave-packet dynamics simulation for \ce{H2} with the goal of finding an initial energy in which the first excited vibrational state is dominant while the energy is close to the experimental excitation energy. 
Details can be found in Appendix \ref{sec:amplitude_dependence}.

We then assess the correctness of our simulation by comparing the calculated spectrum with the experimentally determined frequencies of each mode. 
For each simulation, we record positions and velocities of the atoms as well as the total electric dipole moment, allowing us to calculate velocity power spectra and infrared (IR) spectra, respectively.
Spectral lines are extracted from the calculated spectra by fitting a Gaussian to each peak in the spectrum and taking its centroid as the final frequency.

\vspace{1em}

\section{Results and Discussion} \label{sec:Results}

\subsection{The Hydrogen molecule}

We begin the evaluation of our algorithm with the hydrogen molecule, as it is one of the simplest molecular systems while possessing a vibrational mode that is relatively challenging to predict correctly using both purely electronic and NEO methods, especially while restricted to a double-zeta basis set \cite{Culpitt2025, Wolniewicz1995, CCCBDB2022, Tao2021}.

\begin{table}[!ht]
    \centering
    \caption{The relative errors of the simulated normal mode frequencies for the hydrogen molecule. Frequencies were determined by a Gaussian fit on the velocity power spectrum. The frequencies of the harmonic Hessian approximation at the DFT-level were computed with PBE0 exchange-density functional and the 6-31G basis set. All values are given in percent.}
    \label{tab:h2_freq}
    \begin{tabular}{@{}lcc@{}}
    \toprule
    Method              & \makecell{Purely\\electronic} & NEO \\ \midrule
    UCCSD               & 4.6      & -5.0      \\
    Real Amplitudes     & 10.8      & 0.6      \\
    tUPS                & 5.4      & 0.3      \\
    Hessian (FCI)       & 6.7      & -4.6      \\
    Hessian (DFT)       & 1.8      &           \\\bottomrule
    \end{tabular}
\end{table}

Table \ref{tab:h2_freq} shows the relative errors with respect to the experimental value given in Ref.~\cite{Huber1979} for the frequencies determined by our molecular dynamics quantum algorithm with the UCCSD, real amplitudes and tUPS ansatz in the purely electronic and the NEO case, respectively. We recall that the real amplitudes and tUPS ansatz generally yield shallower circuits with fewer parameters and CNOT gates, making them more suitable for near-term quantum devices with limited coherence times and connectivity. Additionally, standard harmonic Hessian approximations are displayed.

The magnitude of the relative errors for the UCCSD ansatz do not differ significantly between the purely electronic and the NEO case.
It is of note, however, that the NEO method result underestimates the frequency while the purely electronic method overestimates it. 
This aligns with observations in literature \cite{Tao2021, Culpitt2025}, indicating that a systematic underestimation of frequencies is to be expected when employing NEO.
The large shift between the purely electronic and the NEO frequency could also be partially attributed to the fact that the NEO method was not designed to treat all-quantum molecules \cite{Pavoevi2020}.

In the case of the hardware efficient ansatzes, the tUPS ansatz achieves half of the error of the real amplitudes ansatz for both the purely electronic and the NEO calculation.
The NEO results for the hardware efficient ansatzes have an error below one percent.
We argue this is due to fortunate cancelling of errors, where the general overestimation of frequencies by simpler ansatzes is counteracted by the tendency of NEO calculations to underestimate frequencies. 

\subsection{Water}
\begin{figure}
\centering
\includegraphics[width=\columnwidth]{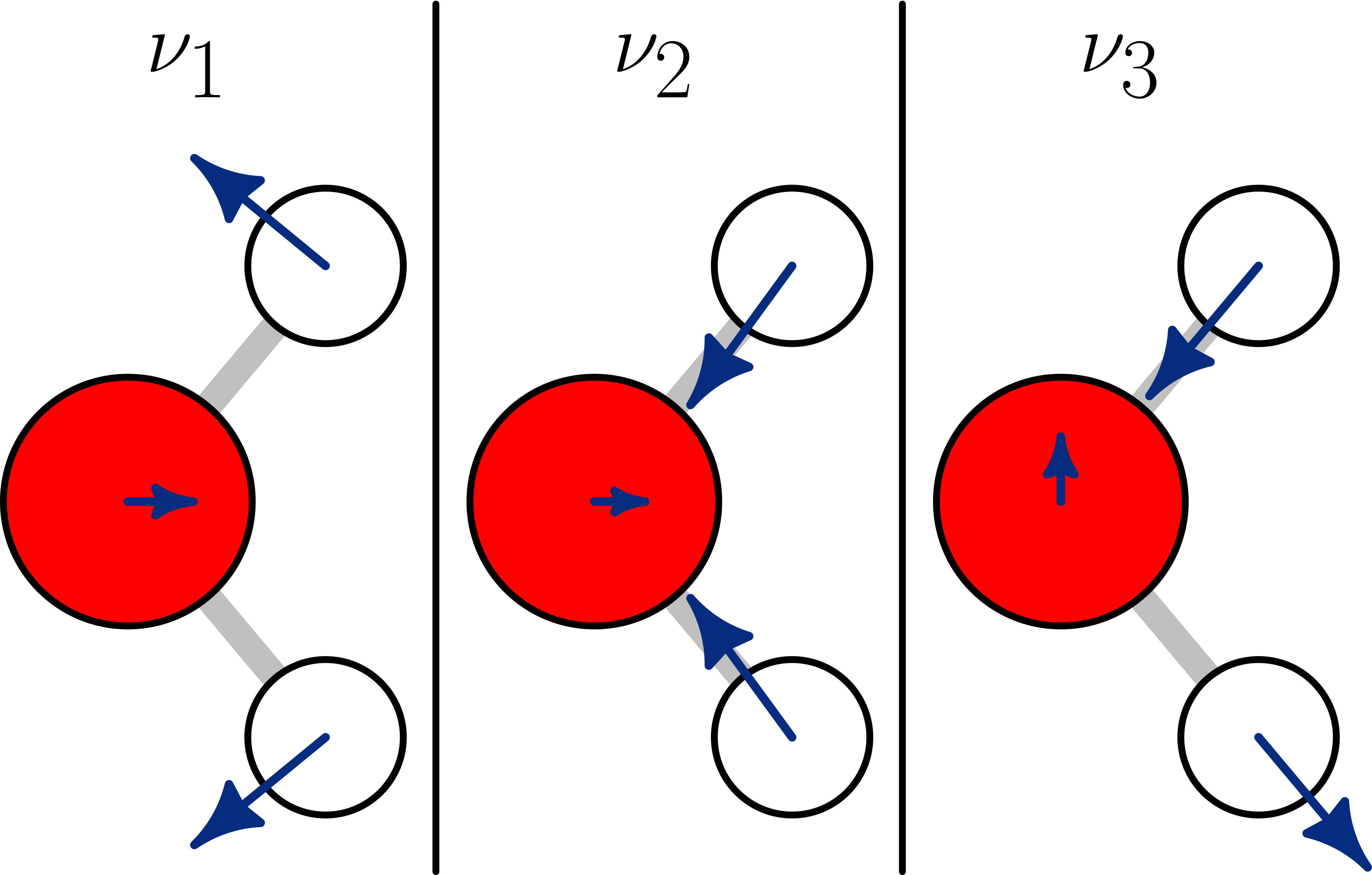}
\caption{The normal modes of water. From left to right: Bending, symmetric stretching and antisymmetric stretching.}
\label{fig:normal_modes_water}
\end{figure}

We further apply our algorithm to explore other chemically more interesting systems, starting with water as a prototypical molecule for which a wealth of experimental benchmark data is available.
The three normal modes of water comprise a bending mode and two stretching modes. 
These modes are displayed in Fig.~\ref{fig:normal_modes_water}.
In the bending mode, the angle between the two \ce{OH} bonds varies, while for the stretching modes, the \ce{OH} lengths vary in phase or in antiphase, respectively.

\begin{figure}
    \centering
    \includegraphics[width=\linewidth]{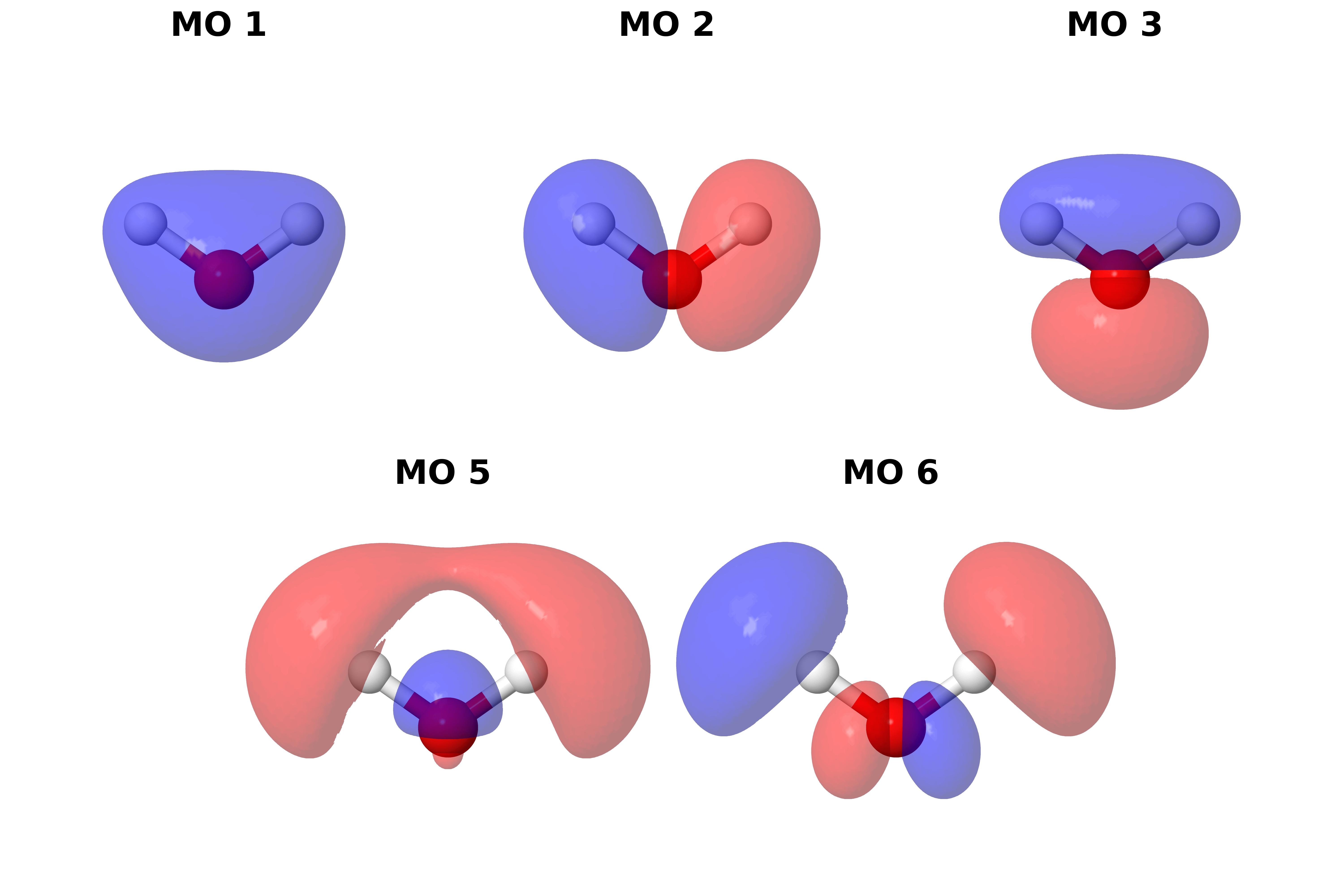}
    \caption{Active electronic orbitals for the water molecule.}
    \label{fig:h2o_active_orbitals}
\end{figure}

The nuclear basis set for the simulations with the water molecule is DZSNB. 
As the normal modes of water do not involve rearrangement of atoms or bonds, there is no clear reaction path that could be used for the second and third selection criterion in Sec.~\ref{sec:ActiveSpaceSelection}. 
Instead, our active space selection rules reduce to choosing the orbitals with the highest single-orbital entropies.
This results in an electronic active space of [1,2,3,5,6], as shown in Figure \ref{fig:h2o_active_orbitals}. 
All nuclear molecular orbitals were treated as active orbitals.
The relative errors of the simulated frequencies from the MD simulations are listed in Table \ref{tab:h2o_freq}, where we used the values from Ref.~\cite{Shimanouchi1972} as experimental reference.

\begin{table*}[!ht]
    \centering
    \caption{Relative errors in percent w.r.t.\ predicted frequencies for the water molecule. The frequencies for the harmonic approximation at the (NEO)CASCI level were computed using the active space presented in Fig. \ref{fig:h2o_active_orbitals}. The frequencies of the harmonic Hessian approximation at the DFT-level were computed with PBE0 exchange-density functional and the 6-31G basis set.
}
    \label{tab:h2o_freq}
    \begin{tabular}{@{}lcccccc@{}}
    \toprule
    Method              & \multicolumn{2}{c}{Mode 1} & \multicolumn{2}{c}{Mode 2} & \multicolumn{2}{c}{Mode 3}  \\
                        & Purely electronic & NEO & Purely electronic & NEO & Purely electronic & NEO \\
    \midrule
    UCCSD               & 6.8 & -0.9   & 5.3 & -1.0   & 2.8 & -1.9 \\
    Real Amplitudes     & 7.6 & 0.0   & 8.7 & 5.8   & 10.6 & 5.9 \\
    tUPS                & 8.0 & 0.1   & 8.4 & 4.8   & 9.7 & 5.8 \\
    Hessian (CASCI)     & 7.1 & 0.3   & 1.8 & -3.4   & 4.2 & -0.7 \\
    Hessian (DFT)       & -2.9 &        & -4.0 &        & -2.3      \\ \bottomrule
    \end{tabular}
\end{table*}

For water, the NEO method achieves better errors than the purely electronic method for all normal modes.
As previously noted for the hydrogen molecule, the systematic underestimation of frequencies by the NEO method can be observed for water as well.

The error of the NEO method is smaller than in the hydrogen simulation, while the purely electronic error is worse for mode 1 and mode 2.
Since the water bending mode (mode 1), where two protons come close to each other, involves a similar interaction between quantum nuclei as does the bond stretching in hydrogen, we pose three hypotheses concerning the performance of the NEO molecular dynamics:

\begin{enumerate}
    \item Direct interaction between quantum nuclei (without intermediary classical nuclei) are hard to simulate.
    \item High-frequency modes are hard to simulate.
    \item Performance issues might stem from relatively small basis set and active spaces. 
\end{enumerate}

Identifying the cause of the accuracy difference between the hydrogen and water simulations for NEO would require a comprehensive study of a wider variety of molecules as well as larger basis sets and active spaces. 
This is a formidable task which is beyond the scope of the current work.

\subsection{Zundel ion}
Finally, we simulate the Zundel ion with our algorithm. 
This is both a prototypical system for studying proton transfer in water \cite{Tuckerman1994, Agmon1995, Tuckerman1995} and displays a behavioural and spectral richness to pose a challenging test case for any MD simulation.

We selected three promising electronic active spaces for the Zundel ion using the approach described in chapter \ref{sec:ActiveSpaceSelection}, namely [7,8,9,10,11] (6e, 5o), [5,8,10,11,14](4e, 5o) and [5,8,14,23,25](4e,5o).
The electronic orbitals that comprise these active spaces are shown in Fig.~\ref{fig:active_orbitals} in the Appendix.

\subsubsection{Potential energy surfaces}

The Zundel cation consists of two water molecules with a proton situated between the two oxygen atoms.
As the proton transfer within the Zundel ion is of special interest, we first investigated the effect of the NEO method on the predicted energy barrier of the proton transfer process.

In order to sample the effective potential energy surface of the proton transfer reaction, we performed a relaxed scan of the proton transfer path using the purely electronic approach. 
To estimate the impact of nuclear quantum effects, we repeated the scan while treating the central proton as a quantum nucleus.
We executed this relaxed scan for each candidate active space.
During optimization, we constrained the distance between the oxygen atoms to be \SI{2.5}{\text{\AA}}. This distance coincides with experimental expectation value of Zundel cations in an aqueous environment \cite{Triolo1975}.
In this constrained geometry, there exist two ground state geometries, where the central proton sits closer to either one of the oxygen atoms.
The start and end points of the relaxed scans were chosen to be the two ground state geometries, respectively and each scan consisted of a total of 40 steps.

\begin{figure}
    \centering
    \includegraphics[width=\linewidth]{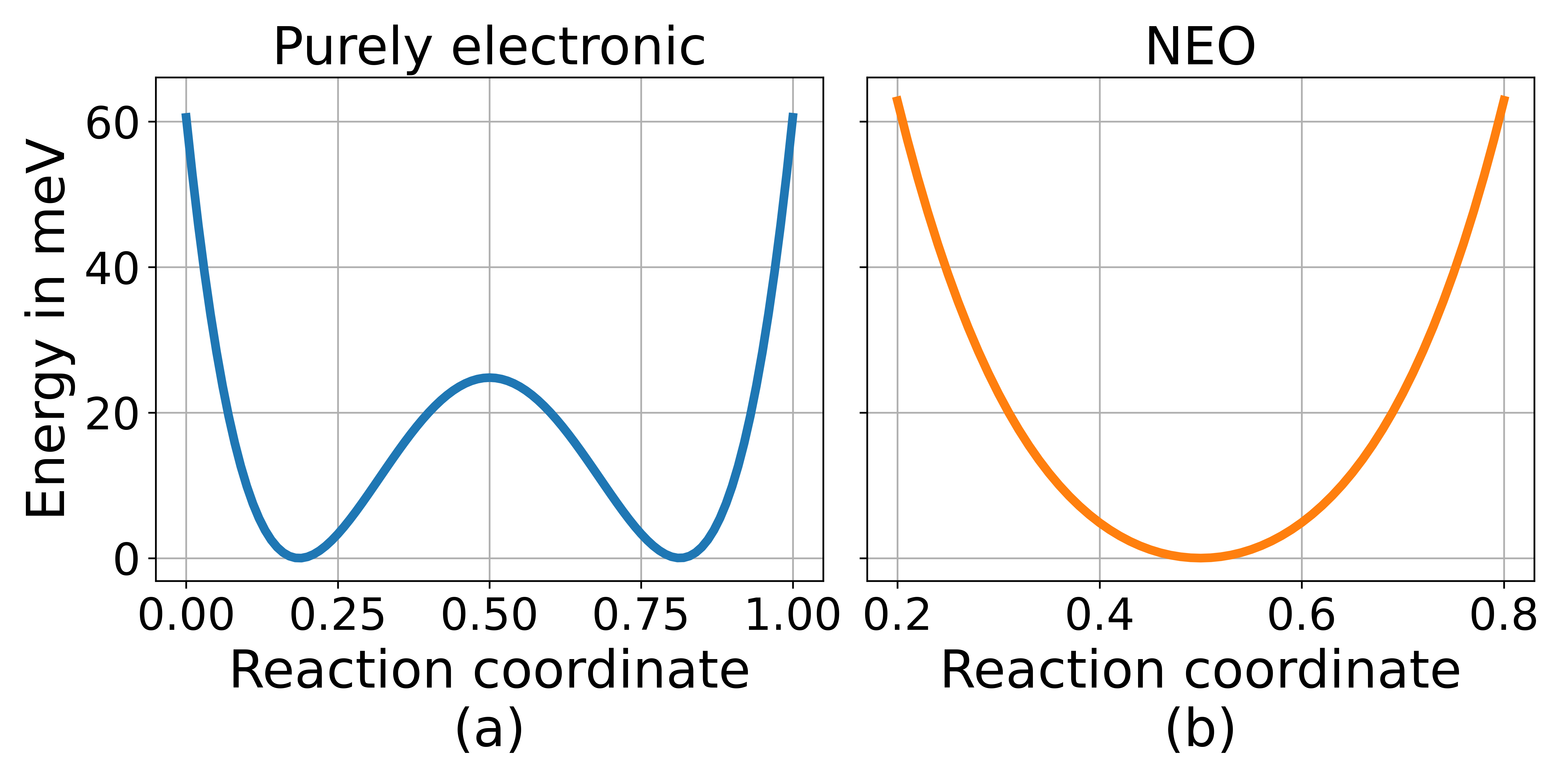}
    \caption{Potential energy along the proton transfer reaction coordinate, computed using the active space [5,8,14,23,25]. (a) Computed with ab-initio electronic structure methods. (b) Computed with the described ab-initio nuclear-electronic structure method.}
    \label{fig:pes_comparison}
\end{figure}

We show representative effective potential energy surfaces for the active space [5,8,14,23,25] in Fig.~\ref{fig:pes_comparison} as we find no qualitative difference between the results for the different active spaces.
In the purely electronic cases, we determined barrier heights of \SI{43.3}{meV} for [7,8,9,10,11], \SI{24.8}{meV} for [5,8,14,23,25] and \SI{12.4}{meV} for [5,8,10,11,14].
In comparison to the thermal energy $k_{\mathrm{B}}T$ at room temperature $T=\SI{298}{K}$, the barrier heights are $1.7\,k_{\mathrm{B}}T$, $1.0\,k_{\mathrm{B}}T$ and $0.5\,k_{\mathrm{B}}T$, respectively.
Other theoretical works predict a barrier height close to $k_{\mathrm{B}}T$ \cite{Lobaugh1996, Pavese1997}, making [5,8,14,23,25] the best performing candidate active space for the prediction of the proton transfer barrier.

For all three investigated active spaces, there is a significant change in character in the proton transfer energy surface upon inclusion of nuclear quantum effects, changing from a double-well to a single-well potential. 
As discussed by Tuckerman~\cite{Tuckerman1997}, this indicates that the vibrational groundstate of the central proton lies above the purely electronic barrier. 
Consequently, the zero-point motion of the shared proton exerts a notable influence even in room-temperature systems.

\subsubsection{Dynamics}

\begin{figure}
    \centering
    \includegraphics[width=\linewidth]{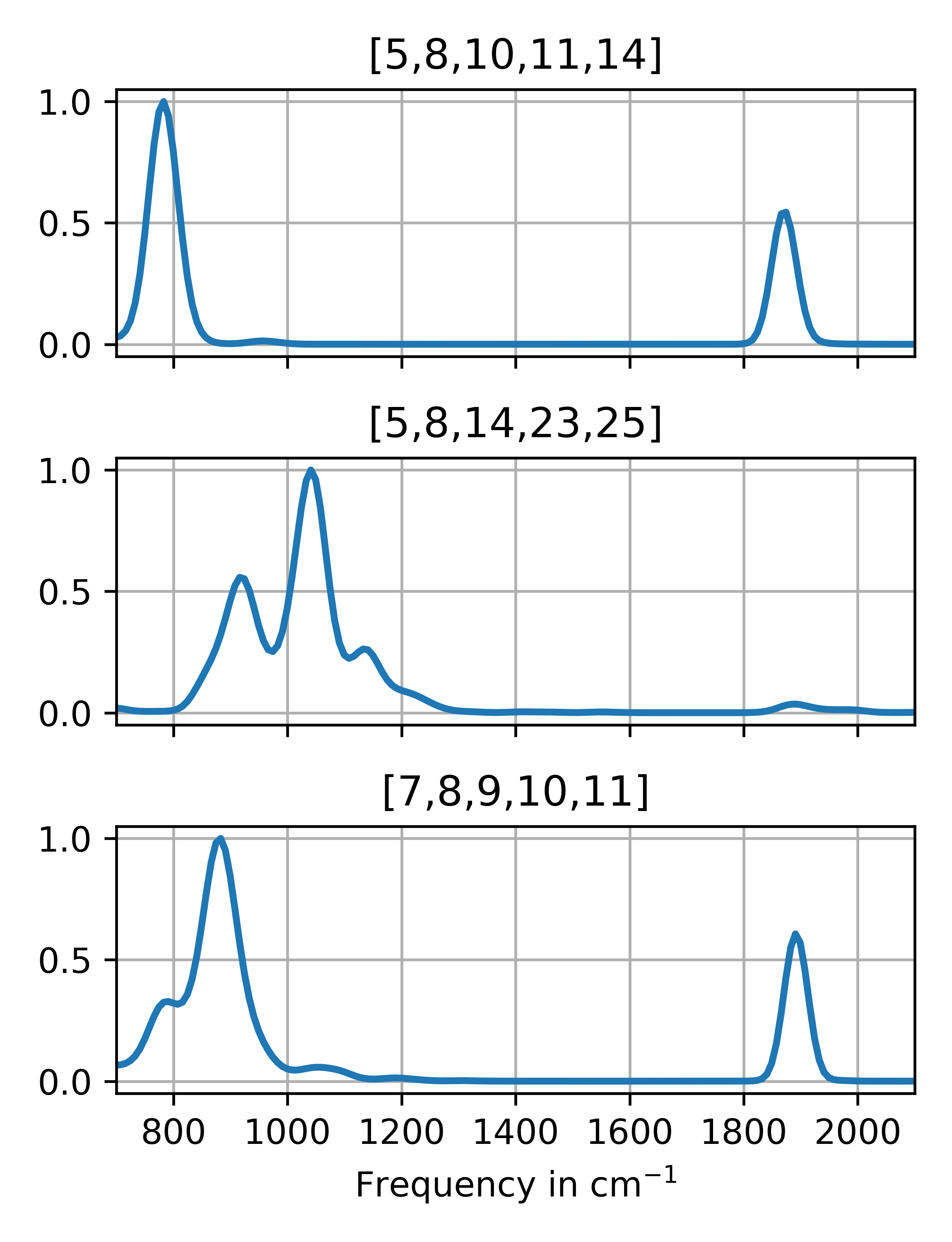}
    \caption{The simulated spectra for the three candidate active spaces using the purely electronic method with the UCCSD ansatz.}
    \label{fig:spectra_zundel_conv}
\end{figure}

We first simulated the proton transfer mode for each candidate active space using the purely electronic method.
The results are shown in Fig.~\ref{fig:spectra_zundel_conv}. 
These results demonstrate a strong sensitivity of the spectral features to the choice of active space, underscoring the critical importance of a careful and systematic active space selection for the accurate description of proton transfer dynamics. 
In particular, they highlight that active orbitals must be chosen with physical and chemical insight rather than by default rules such as selecting orbitals around the Fermi level, which is a common practice in quantum chemistry simulations on quantum computers. 
Our findings suggest that such a simplistic approach may fail to capture the essential electronic correlations and nuclear quantum effects, potentially leading to unphysical results.

\begin{figure}
    \centering
    \includegraphics[width=\linewidth]{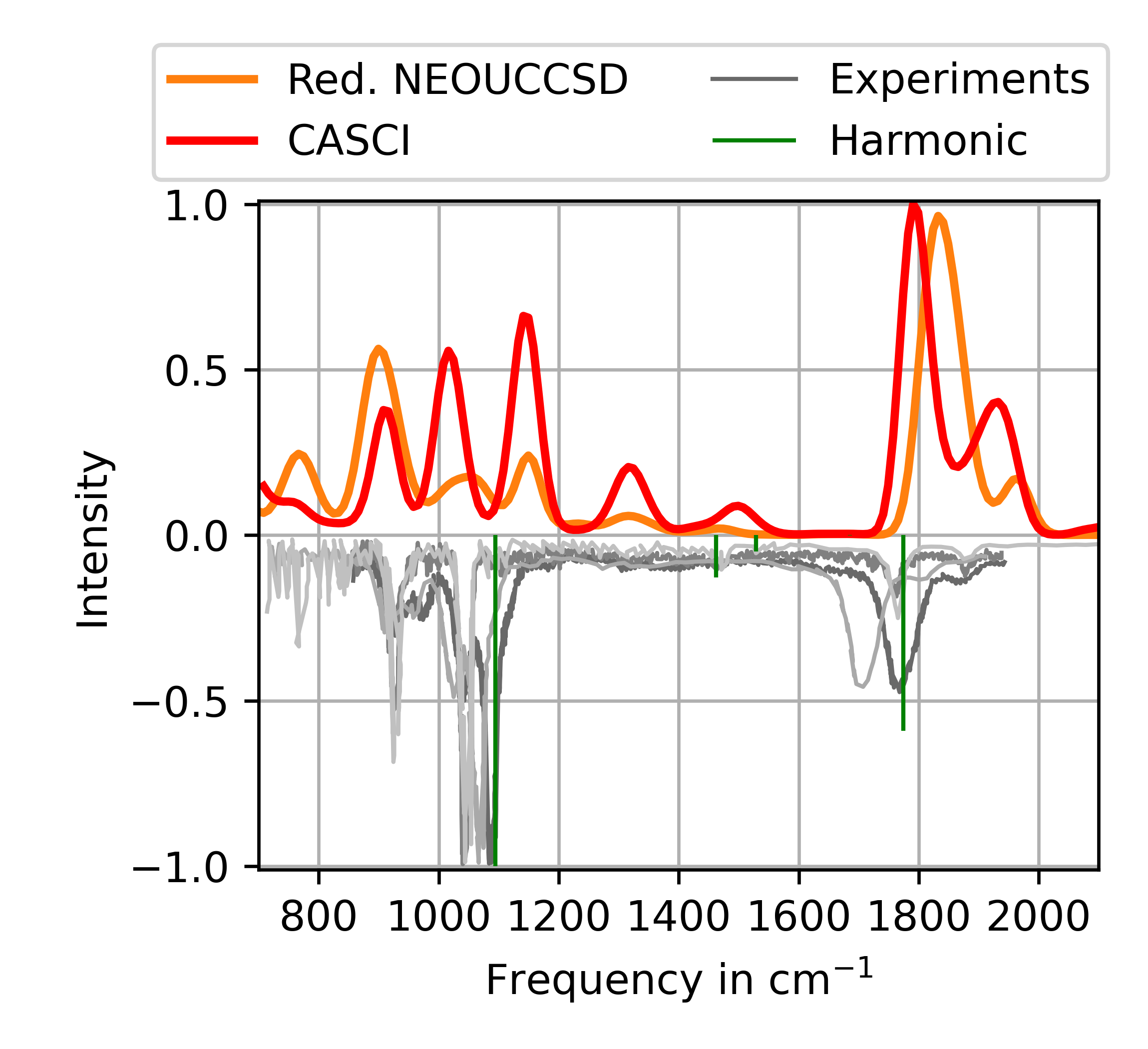}
    \caption{Comparison between the simulated NEO spectrum with $s$- and $p_z$-type nuclear orbitals, normal modes obtained via DFT(PBE0, def2-TZVP) and experimentally obtained predissociation spectra of Zundel ions tagged with Neon and Argon~\cite{Hammer2005, McCunn2007}.}
    \label{fig:zundel_experimental}
\end{figure}

Experimental spectra from Ref.~\cite{Hammer2005, McCunn2007} of the Zundel ion in vacuum are shown in Fig.~\ref{fig:zundel_experimental}. 
The most characteristic signal of the spectrum is the doublet centered around \SI{1000}{\per \centi \metre}. 
Theoretical analyses \cite{Vendrell2007, Vendrell2009, Larsson2022} suggest that this doublet is due to anharmonic coupling between the proton transfer mode and a superposition of the wagging modes of the outer water molecules and the water-water stretch mode. 
At around \SI{1750}{\per \centi \metre}, there is a further notable peak due to bending motions of the outer water moieties.

In comparison to these experimental spectra the active space [5,8,14,23,25] exhibits the most promising features in the region of \SI{900}{\per \centi \metre} -- \SI{1300}{\per \centi \metre}. 
While the spectral features in this simulation look very promising, a visual inspection of the simulated trajectory yields that the proton transfer mode only couples with the water-water stretch mode and not with the water wagging modes.
Based on its good results predicting both the proton transfer barrier and the characteristic doublet spectral signature, we continue with the electronic active space [5,8,14,23,25] for the NEO simulations.

\begin{figure}
    \centering
    \includegraphics[width=\linewidth]{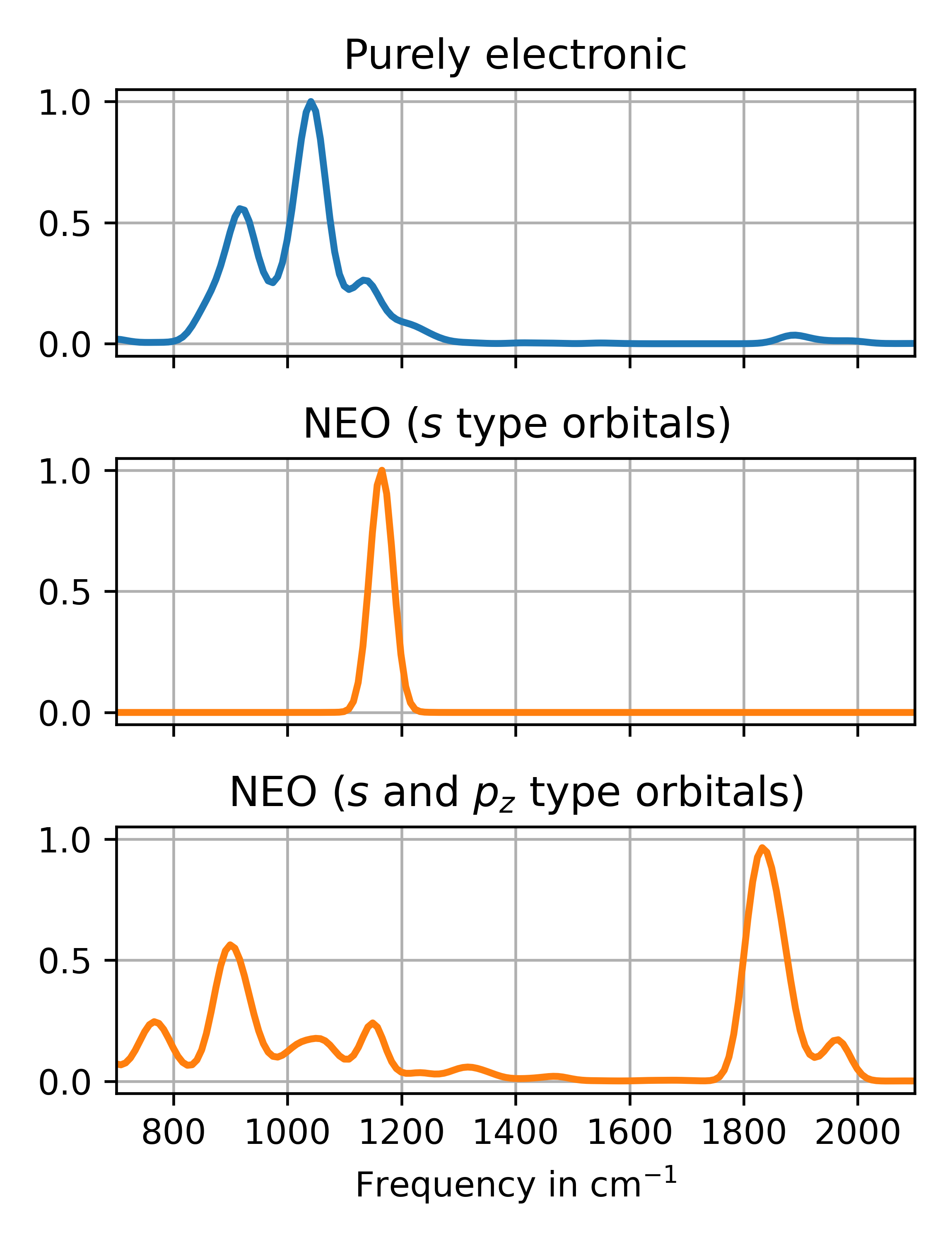}
    \caption{Simulated spectra of the Zundel ion for the electronic active space [5,8,14,23,25]. Top: Purely electronic. Middle: NEO with DZSNB basis set calculated with the full NEOUCCSD ansatz. Bottom: NEO with DZSNB and $p_z$ type nuclear orbitals calculated with the reduced NEOUCCSD ansatz.}
    \label{fig:spectrum_zundel_dzspnb}
\end{figure}

Fig.~\ref{fig:spectrum_zundel_dzspnb} shows a comparison between the spectra for the purely electronic and the NEO simulations with DZSNB and DZSPNB nuclear basis sets. 
While all calculations so far could be performed using the (NEO)UCCSD ansatz, the larger active space of the $p_z$-type NEO Zundel calculations coupled with the small energy gap of \SI{1}{mHa} between the groundstate and the first excited state poses a hard optimization problem that cannot be solved consistently with VQE.
To mitigate this problem, we reduced the NEOUCCSD ansatz for the $p_z$-type simulation by including only the 17 most important excitations instead of the full set of 63 excitations.
The importance of each excitation was determined by repeatedly performing a VQE for the Zundel ion along an exemplary trajectory and scoring each excitation by the sum of its mean parameter value and variability during the trajectory.
We will refer to this ansatz as the "reduced NEOUCCSD" ansatz. Further details can be found in Appendix~\ref{sec:reduced_neouccsd}.

This approach bears similarity to the ADAPT-VQE technique \cite{Grimsley2019} in that we also select a set of one- and two-body operators from a given operator pool in order to construct an efficient ansatz circuit. 
Rather than selecting operators based on a gradient criterion like in ADAPT-VQE, we select excitations based on the mean variability and mean absolute value of their corresponding optimal parameters on multiple representative geometries. 
Our approach thus obtains an efficient ansatz that is expected to perform well for all geometries of the anticipated trajectory, while ADAPT-VQE constructs ansatzes that may be uniquely adapted to only subsets of these geometries.

The $s$-type NEO simulation has a lone peak at \SI{1120}{\per \centi \metre}, representing the proton transfer mode, and possesses no other notable structure.
The $p_z$-type NEO simulation shows peaks at \SI{770}{\per \centi \metre}, \SI{900}{\per \centi \metre}, \SI{1050}{\per \centi \metre}, \SI{1150}{\per \centi \metre}, \SI{1320}{\per \centi \metre}, \SI{1840}{\per \centi \metre}, \SI{1970}{\per \centi \metre}, exhibiting much more structure than the $s$-type simulations.
The additional structure in the $p_z$-type NEO simulation, including the desired spectral doublet, is likely due to the nuclear $p_z$ orbitals increasing the effective nuclear-electronic interaction range facilitating mode coupling.
This is further supported by another visual inspection of the trajectory generated by the $p_z$-type NEO MD, where coupling between the proton transfer mode and both the water wagging and water-water stretch mode is visible and thus improving upon its purely electronic counterpart displayed in the upper panel in Fig.~\ref{fig:spectrum_zundel_dzspnb}.
Similarly, the NEO $s$-type simulation displays a complete lack of mode coupling, as only the initially excited proton transfer mode is visible in the spectrum and any other modes are staying almost completely suppressed.

In Fig.~\ref{fig:zundel_experimental} we compare the simulated NEO spectrum with $s$ and $p_z$-type nuclear orbitals to several experimental spectra \cite{Hammer2005, McCunn2007} and harmonic frequencies calculated with DFT. We note that we focus only on the peak frequencies, rather than their intensities. 
Accurately predicting intensities would likely require electronic basis sets with diffuse and polarization functions \cite{Galabov2002, Silva2012, Krishnadas2026}, which would grow the active space and thus the computational cost of simulating a sufficiently large quantum computer beyond what is tractable.

 Experimentally, the proton transfer doublet consists of a lower peak at \SI{930(10)}{\per \centi \metre} and a higher-frequency peak at \SI{1062(38)}{\per \centi \metre}.

 The NEOCASCI spectrum has peaks at \SI{750}{\per \centi \metre}, \SI{915}{\per \centi \metre}, \SI{1022}{\per \centi \metre}, \SI{1150}{\per \centi \metre}, \SI{1323}{\per \centi \metre}, \SI{1500}{\per \centi \metre}, \SI{1800}{\per \centi \metre} and \SI{1930}{\per \centi \metre}. 
Since the NEOCASCI calculation represents the exact solution within the active space, any deviations from it by the reduced NEOUCCSD dynamics can be interpreted as the error introduced by the quantum simulation itself.
Here, the maximum error per peak is less than \SI{40}{\per \centi \metre}.
The main contribution to the total error is due to the chosen ansatz circuit not being able to exactly represent the ground state.
Additionally, both the purely electronic as well as the DZSPNB nuclear results qualitatively predict the proton transfer doublet and estimate its frequencies with a small error of about \SI{10}{\per \centi \metre}.
 On the other hand, both methods overestimate the frequencies of the outer water stretch mode at around \SI{1800}{\per \centi \metre}, where the purely electronic exhibits a stronger blueshift than the NEO method.

These results demonstrate a strong dependency on the employed electronic active space and nuclear basis set, highlighting the challenge in correctly simulating the proton transfer motion in the Zundel ion.
This difficulty is further emphasized by our simulations of the Zundel ion with the hardware-efficient ansatzes, which are shown in Appendix~\ref{sec:app_more_simulated_spectra}, as the results of the simulations employing the hardware-efficient ansatzes do not match the experimental spectrum well.
As we have demonstrated, this sensitivity can be overcome via a careful choice of active space and nuclear basis set

\subsection{Quantum Resources}
\begin{table*}
\caption{Parameter count, CNOT gate count, and circuit depth of the different simulations. The variable $L$ stands for the number of layers (repetitions) in the hardware-efficient ansatzes.}
\label{tab:quant_resources}
\centering
\begin{tabular}{@{}lccccccccc}\toprule
& \multicolumn{3}{c}{\ce{H2}} & \multicolumn{3}{c}{\ce{H2O}} & \multicolumn{3}{c}{\ce{H5O2+}} \\
\cmidrule(lr){2-4}\cmidrule(lr){5-7}\cmidrule(lr){8-10}
 & Params & CNOT & Depth & Params & CNOT & Depth & Params & CNOT & Depth \\\midrule
UCCSD         & 15       & 560    & 936 & 54 & 3284 & 5035 & 54 & 3284 & 4975\\
\makecell[l]{NEO-\\UCCSD}      & 23       & 632    & 936 & 68 & 3564 & 53496 & 61 & 3493 & 5430\\
real Amp.     & $6(L+1)$ & $5L$   & $3L$   & $8(L+1)$  & $7L$  &$3L$& $8(L+1)$ & $7L$&$3L$ \\
\makecell[l]{NEO-\\real Amp.} & $9(L+1)$ & $8L$   & $3L$   & $11(L+1) $& $10L$ &$3L$& $9(L+1)$ & $8L$&$3L$ \\
tUPS          & $9L$     & $144L$ & $312L$ & $12L$ & $208L$ &$333L$& $12L$ & $208L$&$333L$ \\
\makecell[l]{NEO-\\ tUPS}      & $18L$    & $152L$ & $312L$ & $21L$ & $256L$ &$449L$& $15L$ & $172L$&$320L$ \\\bottomrule
\end{tabular}
\end{table*}

From the previous Tables \ref{tab:h2_freq} and \ref{tab:h2o_freq} as well  Fig.~\ref{fig:zundel_spectra_ele_hea} and Fig.~\ref{fig:zundel_spectra_neo_hea} in Appendix~\ref{sec:app_more_simulated_spectra} it can be gathered that the most accurate simulations employed the (NEO)UCCSD ansatz. 
On the other hand, hardware-efficient ansatzes deliver acceptable results for \ce{H2} and \ce{H2O} while requiring significantly less quantum resources (see Table \ref{tab:quant_resources}).
For the Zundel ion, the hardware-efficient ansatzes fail to correctly capture the non-linear couplings between the different normal modes.

Despite this, the real amplitudes ansatz in particular seems like a promising candidate for running our algorithm on NISQ hardware, as its shallow depth and nearest-neighbor topology make it well suited for current quantum devices. 
We expect the results of the real amplitudes ansatz to be systematically improvable by enlarging the active space. 
While a larger active space requires more qubits, the real amplitudes ansatz demonstrates favorable scaling of its depth with respect to the number of qubits.

\vspace{1em}

\section{Conclusion}

\label{sec:Summary}
We implemented a molecular dynamics simulation which evaluates molecular energies and forces on a quantum computer, while integrating the classical equations of motions for the nuclei on a classical computer.
We incorporated nuclear quantum effects into our simulations by employing the NEO framework.
Using this method, we simulated the vibrational modes for \ce{H2}, \ce{H2O} and the Zundel ion \ce{H5O2+}.

We observed that employing NEO improves normal mode simulations for water and the Zundel ion, while the \ce{H2} stretch mode frequency gets overcorrected.
While purely $s$-type nuclear orbitals are sufficient for the water molecule, the Zundel ion additionally required nuclear $p$-type orbitals for adequate replication of anharmonic mode couplings.
This demonstrates the importance of including nuclear quantum effects for accurately modelling the proton transfer in the Zundel ion.

Our hybrid MD algorithm produced the highest-quality results in conjunction with the UCCSD ansatz, with an average error of less than \SI{5}{\percent} while applying a relatively small active space of five to seven orbitals, or around ten qubits.
Hardware-efficient ansatzes generally perform very closely to their UCCSD counterparts at significantly lower quantum computing resource requirements.
 
In this work, we worked within the canonical orbital basis.
We expect further improvements in either quantum resource usage or alternatively accuracy by changing to a natural orbital basis, as natural orbitals provide faster CI convergence than canonical orbitals do.
Our experiments with the Zundel ion demonstrated that the correct determination of active electronic and nuclear orbitals, irrespective of the concrete chosen orbital basis, is essential for achieving reliably accurate predictions.

\vspace{1em}

\section*{Acknowledgments}
 
This project was made possible
by the DLR Quantum Computing Initiative and the
Federal Ministry for Economic Affairs and Climate
Action; \href{https://qci.dlr.de/basiq/}{qci.dlr.de/projects/basiq}.  For the emulations,
the authors gratefully acknowledge the scientific support
and HPC resources provided by the German Aerospace
Center (DLR). The HPC system CARA is partially
funded by ”Saxon State Ministry for Economic Affairs,
Labour and Transport” and ”Federal Ministry for
Economic Affairs and Climate Action”.
This work contributes to the research performed at CELEST (Center for Electrochemical Energy Storage Ulm-Karlsruhe).

\vspace{1em}

\bibliographystyle{aip_mod}
\bibliography{bibliography}

\vspace{1cm}

\appendix

\section{Candidate electronic molecular orbitals for the Zundel Ion}
\label{sec:app_active_orbitals_zundel}
\begin{figure}[ht]
    \centering
    \includegraphics[width=\linewidth]{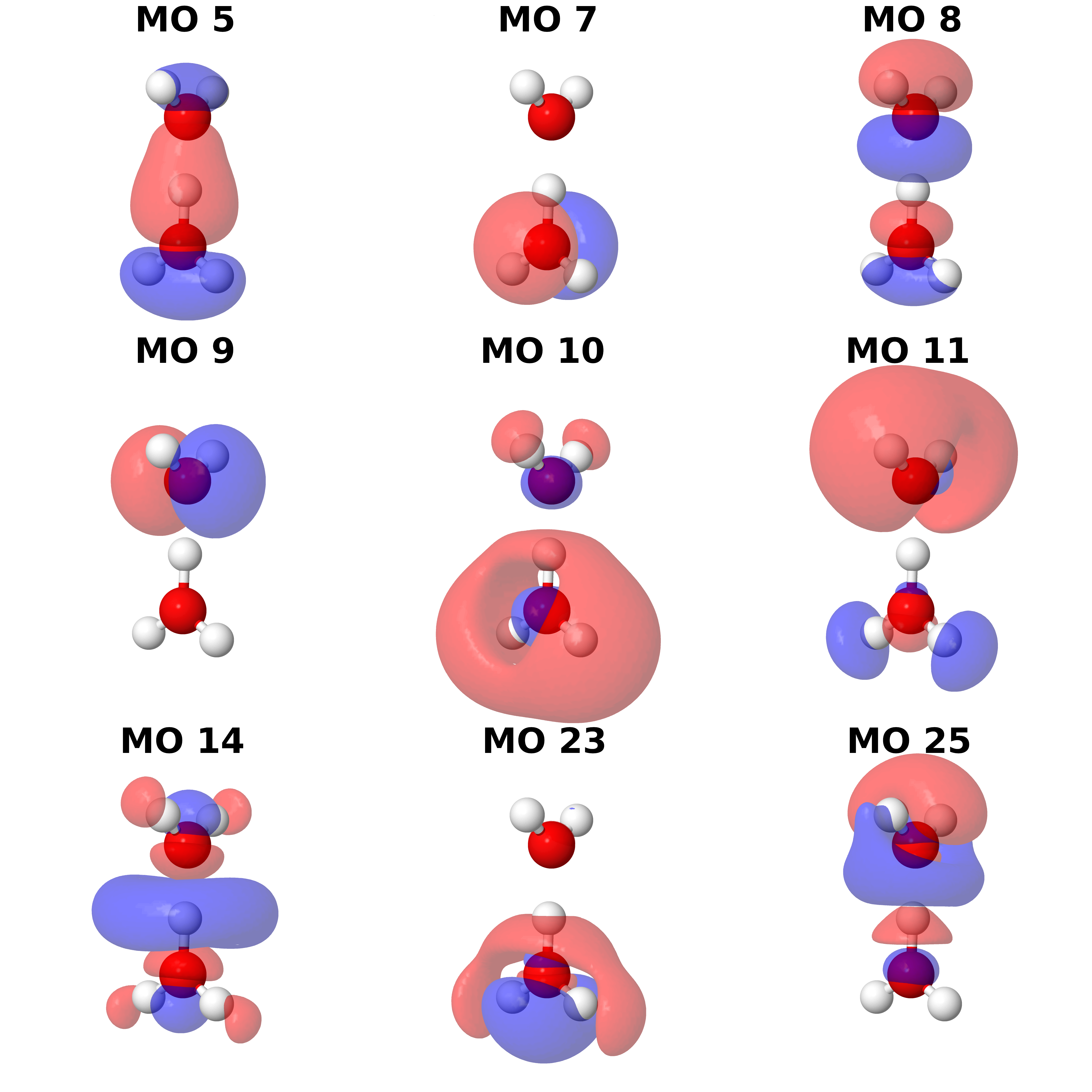}
    \caption{The electronic molecular orbitals for the reference geometry of the Zundel ion which compose the investigated electronic active spaces [7,8,9,10,11], [5,8,10,11,14] and [5,8,14,23,25].}
    \label{fig:active_orbitals}
\end{figure}
Fig.~\ref{fig:active_orbitals} shows the collection of electronic MOs of the Zundel ion that were considered for constructing an active electronic space.

\vspace{1em}

\section{Analytical Gradient Operators}
\label{sec:gradient_operators}

The analytical expression for $dH_\text{NEO}/dR$ in the MO basis as given by Webb et al. \cite{Webb2002} is 
\begin{align}
    \frac{dH_\text{NEO}}{dR_{Ik}} &= \sum_{pq} T^\text{e}_{pq} a^\dagger_p a_q + \sum_{P Q} T^\text{n}_{P Q} a^\dagger_{P} a_{Q} \nonumber \\
    &+ \sum_{pqrs} V^\text{e}_{pqrs} a^\dagger_{p} a^\dagger_{r} a_s a_q  \nonumber \\
    &+ \sum_{P Q R S} V^\text{n}_{P Q R S} a^\dagger_{P} a^\dagger_{R} a_{S} a_{Q}\nonumber\\
    &+ \sum_{pq}\sum_{P Q} V^\text{en}_{pqP Q} a^\dagger_{p} a^\dagger_{P} a_Q a_{q} \,, \label{eq:ana_deriv}
\end{align} where the electronic and nuclear single-body derivatives $T^\text{e}_{pq}$ and  $T^\text{n}_{PQ}$ are
\begin{align}
    T^\text{e}_{pq} &= \frac{dh^\text{e}_{pq}}{dR_{Ik}} - \sum_m h^\text{e}_{mq} \frac{dS^\text{e}_{mp}}{dR_{Ik}} \nonumber \\
    T^\text{n}_{P Q} &= \frac{dh^\text{n}_{P Q}}{dR_{Ik}} - \sum_{M} h^\text{n}_{M Q} \frac{dS^\text{n}_{M P}}{dR_{Ik}} \nonumber \,,
\end{align}
and the two-body derivatives are
\begin{align}
    V^\text{e}_{pqrs} &= \frac{dg^\text{e}_{pqrs}}{dR_{Ik}} - 2 \sum_t g^\text{e}_{tqrs} \frac{dS^\text{e}_{tp}}{dR_{Ik}} \nonumber \\
    V^\text{n}_{P Q R S} &= \frac{dg^\text{n}_{P Q R S}}{dR_{Ik}} - 2 \sum_{T} g^\text{n}_{T Q R S } \frac{dS^\text{n}_{T P}}{dR_{Ik}} \nonumber \\
    V^\text{en}_{pqP Q} &= -\frac{dg^\text{en}_{pqP Q}}{dR_{Ik}} + \sum_k g^\text{en}_{kqP Q} \frac{dS^\text{e}_{kp}}{dR_{Ik}} \nonumber\\
    \label{eq:gradient_short_form}
    &+ \sum_{K} g^\text{en}_{pqK Q} \frac{dS^\text{n}_{K P}}{dR_{Ik}} \,.
\end{align}

Here, $S^\text{e}$ and $S^\text{n}$ are the electronic and nuclear overlap integrals.
The derivatives $\frac{dS_{mp}}{dR_{Ik}}$,  $\frac{dh_{pq}}{dR_{Ik}}$ and $\frac{dg_{pqrs}}{dR_{Ik}}$ are so-called "skeleton derivatives", which represent the change of atomic orbital (AO) integrals in the MO basis, for example
$$
\frac{dS_{pq}}{dR_{Ik}} = \sum_{\mu \nu} C_{\mu p} C_{\nu q} \frac{dS_{\mu \nu}}{dR_{Ik}} \,.
$$ 
These AO integral derivatives can be easily obtained from most quantum chemistry software packages.
There is another more general expression given in Ref.~\cite{OBrien2022} that computes the contractions with the skeleton derivatives of the overlap integrals by involving all indices of the one- and two-body integrals:

\begin{align}
    T_{pq} &= \frac{dh_{pq}}{dR_{Ik}} - \frac{1}{2}\sum_m \left( h_{mq} \frac{dS_{pm}}{dR_{Ik}} + h_{pm} \frac{dS_{mq}}{dR_{Ik}}\right) \nonumber \\
    V_{pqrs} &= \frac{dg_{pqrs}}{dR_{Ik}} - \frac{1}{2} \sum_t \bigg( g_{tqrs} \frac{dS_{pt}}{dR_{Ik}}  + g_{ptrs} \frac{dS_{tq}}{dR_{Ik}} \nonumber \\
    &+ g_{pqts} \frac{dS_{rt}}{dR_{Ik}} + g_{pqrt} \frac{dS_{ts}}{dR_{Ik}}\bigg)  \,. \label{eq:force_long} 
\end{align}

These two variants are equivalent when they are evaluated on the same density matrix (see Appendix \ref{sec:compare_force_op_no_embedding}). However, if they are evaluated on two distinct density matrices, for example due to an active space method, the results may not coincide. We address this in Sec.~\ref{sec:active_space_transformation} and App.~\ref{sec:compare_force_ops} below.

\vspace{1em}

\section{Comparison of the two formulations of the force operators}
\label{sec:compare_force_ops}

For the sake of clarity, in the following we will use the Einstein sum notation and write $\frac{dS_{pq}}{dR_{Ik}}$ as $\Dot{S}_{pq}$.

\subsection{No embedding}
\label{sec:compare_force_op_no_embedding}

First, we compare the two formulations of the one-body terms in (\ref{eq:gradient_short_form}) and (\ref{eq:force_long}), which we will call $T^{\text{short}}_{pq}$ and $T^{\text{long}}_{pq}$, respectively. 
We first show that their contractions with the same density matrix $\gamma_{pq} = \langle a^\dagger_p a_q\rangle$ are equal. 
Starting from (\ref{eq:force_long}):
\begin{align*}
    &\gamma_{pq} T^{\text{long}}_{pq}\\
    &=\gamma_{pq}\frac{dh_{pq}}{dR_{Ik}} - \frac{1}{2} \gamma_{pq} h_{mq} \Dot{S}_{pm} - \frac{1}{2} \gamma_{pq} h_{pm} \Dot{S}_{mq}\\
    &=\gamma_{pq}\frac{dh_{pq}}{dR_{Ik}} - \frac{1}{2} \gamma_{pq} h_{mq} \Dot{S}_{pm} - \frac{1}{2} \gamma_{qp} h_{qm} \Dot{S}_{mp}\\
    &=\gamma_{pq}\frac{dh_{pq}}{dR_{Ik}} - \frac{1}{2} \gamma_{pq} h_{mq} \Dot{S}_{mp} - \frac{1}{2} \gamma_{pq} h_{mq} \Dot{S}_{mp}\\
    &=\gamma_{pq}\frac{dh_{pq}}{dR_{Ik}} - \gamma_{pq} h_{mq} \Dot{S}_{mp}\\
    &= \gamma_{pq} \left(\frac{dh_{pq}}{dR_{Ik}} -  h_{mq} \Dot{S}_{mp}\right)\\
    &= \gamma_{pq} T^{\text{short}}_{pq},
\end{align*}
where we renamed indices in the second step and exploited symmetries in the third step.

Continuing with the two-body terms $V^{\text{short}}_{pqrs}$ and $V^{\text{long}}_{pqrs}$ and the two-body reduced density matrix $\Gamma_{pqrs} = \langle a^\dagger_p a^\dagger_r a_s a_q \rangle$, following the same steps from above: 
\begin{align*}
    &\Gamma_{pqrs} V^{\text{long}}_{pqrs} \\
    &= \Gamma_{pqrs}\frac{dg_{pqrs}}{dR_{Ik}} - \frac{1}{2}\Gamma_{pqrs} g_{tqrs} \Dot{S}_{pt}  - \frac{1}{2}\Gamma_{pqrs} g_{ptrs} \Dot{S}_{tq}
    \\&-\frac{1}{2}\Gamma_{pqrs} g_{pqts} \Dot{S}_{rt} - \frac{1}{2}\Gamma_{pqrs} g_{pqrt} \Dot{S}_{ts}\\
    &= \Gamma_{pqrs}\frac{dg_{pqrs}}{dR_{Ik}} - \frac{1}{2}\Gamma_{pqrs} g_{tqrs} \Dot{S}_{pt}  - \frac{1}{2}\Gamma_{qprs} g_{qtrs} \Dot{S}_{tp}
    \\&-\frac{1}{2}\Gamma_{rspq} g_{rstq} \Dot{S}_{pt} - \frac{1}{2}\Gamma_{rsqp} g_{rsqt} \Dot{S}_{tp} \\
    &= \Gamma_{pqrs}\frac{dg_{pqrs}}{dR_{Ik}} - \frac{1}{2}\Gamma_{pqrs} g_{tqrs} \Dot{S}_{tp}  - \frac{1}{2}\Gamma_{pqrs} g_{tqrs} \Dot{S}_{tp}
    \\&-\frac{1}{2}\Gamma_{pqrs} g_{tqrs} \Dot{S}_{tp} - \frac{1}{2}\Gamma_{pqrs} g_{tqrs} \Dot{S}_{tp} \\
    &= \Gamma_{pqrs}\frac{dg_{pqrs}}{dR_{Ik}} - 2\Gamma_{pqrs} g_{tqrs} \Dot{S}_{tp}\\
    &=\Gamma_{pqrs}\left(\frac{dg_{pqrs}}{dR_{Ik}} - 2 g_{tqrs} \Dot{S}_{tp}\right)\\
    &=\Gamma_{pqrs} V^{\text{short}}_{pqrs} \,.
\end{align*}

This result applies to both the electronic and nuclear one- and two-particle operators. 
It is easy to see that this also applies to the electronic-nuclear interaction terms, because they effectively follow the derivation for the one-particle operators.

\subsection{Hartree-Fock embedding}

In the HF embedding, we obtain an expression for the force by inserting the embedding Hamiltonian (Eq.~\eqref{eq:embedding_Hamiltonian}) into the general form of the Hamiltonian gradient (Eq.~\eqref{eq:ana_deriv}). 
We employ the compact (symmetry-reduced) form of the inactive Fock operator (Eq.~\eqref{eq:inactive_Fock_short}) and restrict to the electronic degrees of freedom for the purpose of this derivation (dropping the superscript e for brevity); the nuclear and electronic-nuclear terms follow analogously. We then have
\begin{align*}
-F &=2 T_{jj} + 2V_{iijj} - V_{ijji} + \gamma_{uv} T_{uv} + 2\gamma_{uv}V_{iiuv} \\
&- \gamma_{uv} V_{ivui} + \frac{1}{2}\Gamma_{uvxy} V_{uvxy}. 
\end{align*}

Above, it was established that the contraction of the long or short form force operators with a single density matrix result in the same values.
Operators that are contracted exclusively with either the Hartree-Fock density matrix $\gamma_{ij}^\text{HF}=\delta_{ij}$ ($2 T_{jj}$ and $2V_{iijj}$) or one of the active density matrices ($\gamma_{uv} T_{uv}$ and $\frac{1}{2}\Gamma_{uvxy} V_{uvxy}$) will therefore numerically contribute the same to the force, independent of whether the long or the short form was used.
This leaves the term $2\gamma_{uv}V_{iiuv} - \gamma_{uv} V_{ivui}$ for investigation.
\begin{align*}
    &\gamma_{uv} V^{\text{long}}_{iiuv}\\
    &= \gamma_{uv}\frac{dg_{iiuv}}{dR_{Ik}} - \frac{1}{2}\gamma_{uv} g_{tiuv} \Dot{S}_{it}  - \frac{1}{2}\gamma_{uv} g_{ituv} \Dot{S}_{ti}
    \\&-\frac{1}{2}\gamma_{uv} g_{iitv} \Dot{S}_{ut} - \frac{1}{2}\gamma_{uv} g_{iiut} \Dot{S}_{tv} \\
    &=\gamma_{uv}\frac{dg_{iiuv}}{dR_{Ik}} - \frac{1}{2}\gamma_{uv} g_{tiuv} \Dot{S}_{it}  - \frac{1}{2}\gamma_{uv} g_{tiuv} \Dot{S}_{ti}
    \\&-\frac{1}{2}\gamma_{uv} g_{iitv} \Dot{S}_{ut} - \frac{1}{2}\gamma_{vu} g_{iivt} \Dot{S}_{tu} \\
    &=\gamma_{uv}\frac{dg_{iiuv}}{dR_{Ik}} - \gamma_{uv} g_{tiuv} \Dot{S}_{it} - \gamma_{uv} g_{iitv} \Dot{S}_{ut} \\
     &=\gamma_{uv}\left(\frac{1}{2}\frac{dg_{iiuv}}{dR_{Ik}} - g_{tiuv} \Dot{S}_{it} + \frac{1}{2}\frac{dg_{uvii}}{dR_{Ik}} - g_{tvii} \Dot{S}_{ut}\right) \\
     &=\gamma_{uv} \left(\frac{1}{2} V^{\text{short}}_{iiuv} + \frac{1}{2} V^{\text{short}}_{uvii}\right)\\
\end{align*}
A similar calculation can be performed for the term $\gamma_{uv} V_{ivui}$.

We therefore conclude that the gradient of the short (symmetry-reduced) Fock operator needs to be corrected by the term $\frac{1}{2}\left(-2V^{\text{short}}_{iiuv} - V^{\text{short}}_{ivui} + 2 V^{\text{short}}_{uvii} + V^{\text{short}}_{viiu} \right)$.

This translates to a correction of $\frac{1}{2}\left(-2g_{iiuv} - g_{ivui} + 2 g_{uvii} + g_{viiu} \right)$ for the conventional Fock matrix, thus arriving at the general form Eq.~\eqref{eq:inactive_Fock_long}.

\vspace{1em}

\section{NEOCPHF}

An alternative way of obtaining nuclear derivatives of the MO coefficients is the Coupled Perturbed Hartree-Fock (CPHF) method \cite{Gerratt1968}.
Here we present the extension of the CPHF method to NEO. 

NEOCPHF requires the AO integrals $S^\text{e}, S^\text{n}, h^\text{e}, g^\text{e}, h^\text{n}, g^\text{n}, g^\text{en}$, the integral derivatives $\frac{dS^\text{e}}{dR_{Ik}}, \frac{dS^\text{n}}{dR_{Ik}}, \frac{dh^\text{e}}{dR_{Ik}}, \frac{dh^\text{n}}{dR_{Ik}}, \frac{dg^\text{e}}{dR_{Ik}}, \frac{dg^\text{n}}{dR_{Ik}}, \frac{dg^\text{en}}{dR_{Ik}}$ with respect to nucleus $i$ along dimension $k$ and the MO energies and coefficients $\varepsilon^\text{e}, \varepsilon^\text{n}, C^\text{e}, C^\text{n}$.

The electronic and nuclear Fock matrices are (cf.\ Eq.~\eqref{eq:inactive_Fock_short} in the main text)
\begin{align}
    F^\text{e}_{pq} &= h^\text{e}_{pq} + \sum_{rs} \gamma^\text{e}_{sr} \left( g^\text{e}_{pqrs} - \frac{1}{2} \cdot g^\text{e}_{psqr} \right) \nonumber\\&- \sum_{R S} \gamma^\text{n}_{S R} g^\text{en}_{pqR S}\\
    F^\text{n}_{P Q} &= h^\text{n}_{P Q} + \sum_{R S} \gamma^\text{n}_{S R} \left(g^\text{n}_{P Q S R} - g^\text{n}_{P S Q R}\right) \nonumber\\&- \sum_{rs} \gamma^\text{e}_{sr} g^\text{en}_{rsP Q},
\end{align}
with $\gamma$ being the NEO-HF one-body reduced density matrices.

Furthermore, let the auxiliary matrices $W^\text{e}, I^\text{e}, W^\text{n}, I^\text{n}$ be
\begin{align}
    W^\text{x}_{ij} &= \begin{cases}
    0, &\text{if $i=j$ or $\varepsilon^\text{x}_i = \varepsilon^\text{x}_j$} \\
    \frac{1}{\varepsilon^\text{x}_i - \varepsilon^\text{x}_j}, &\text{else}
    \end{cases}\\
    I^\text{x}_{ij} &= \begin{cases}
    1, &\text{if $i=j$ or $\varepsilon^\text{x}_i = \varepsilon^\text{x}_j$} \\
    0, &\text{else}
    \end{cases}
\end{align}
for $\text{x}=\text{e},\text{n}$.

Then, start with an initial guess for the MO coefficient perturbation $U^{\text{e}(0)}, U^{\text{n}(0)}$ and form the MO coefficient derivatives $\frac{dC^{\text{e}(0)}}{dR}=C^\text{e} U^{\text{e}(0)T}$ and $\frac{dC^{\text{n}(0)}}{dR}=C^\text{n} U^{\text{n}(0)T}$.

Until convergence, the MO coefficient derivatives are updated according to:
\begin{align}
    U^{(i+1)} &= W \circ \left[ C^T \frac{\partial F^{(i)}}{\partial R} C - C^T \frac{\partial S}{\partial R} C \varepsilon \right] \nonumber \\
    &- I \circ \frac{1}{2} \left[C^T \frac{\partial S}{\partial R} C\right],
\end{align}
where $\circ$ is the element-wise product and the Fock matrix derivatives $\frac{\partial F^{(i)}}{\partial R}$ are formed from the current guesses for the MO derivatives $\frac{dC^{\text{e}(i)}}{dR}$, $\frac{dC^{\text{n}(i)}}{dR}$ and the integral derivatives.

On the basis of DDIIS \cite{Weber2003} for CPHF and NEODIIS \cite{Liu2022} for NEOSCF, we introduce NEODDIIS, which is the DDIIS equivalent for NEOCPHF. 
NEODDIIS extrapolates the electronic and nuclear Fock matrix derivatives $\frac{dF}{dR}$ using the Fock matrix derivatives from previous iterations, thus improving the convergence speed of the NEOCPHF algorithm.
Let $[A,B]_C := ACB - BCA$ and let $n$ be the number of previous iterations that will be used for the extrapolation. Let the error vectors be
\begin{align}
\mathbf{e}^\text{e}_i &= \left[\frac{dF^{\text{e}(i)}}{dR}, S^\text{e}\right]_{\gamma^\text{e}} + \left[F^\text{e}, \frac{dS^\text{e}}{dR}\right]_{\gamma^\text{e}} \nonumber \\&+ \left[F^\text{e},S^\text{e} \right]_{\frac{d\gamma^\text{e}}{dR}} \\
\mathbf{e}^\text{n}_i &= \left[\frac{dF^{\text{n}(i)}}{dR}, S^\text{n}\right]_{\gamma^\text{n}} + \left[F^\text{n}, \frac{dS^\text{n}}{dR}\right]_{\gamma^\text{n}} \nonumber \\&+ \left[F^\text{n},S^\text{n} \right]_{\frac{d\gamma^\text{n}}{dR}}.
\end{align}

Then, let the matrix $B \in \mathbb{R}^{(n+1) \times (n+1)}$ be
\begin{align*}
&B_{ij} =\\ &\begin{cases}
\Tr{\mathbf{e}^\text{e}_i \mathbf{e}^\text{e}_j} + \Tr{\mathbf{e}^\text{n}_i \mathbf{e}^\text{n}_j}&, \text{$i<n+1$ and $j<n+1$} \\
-1&, \text{$i=n+1$ xor $j=n+1$} \\
0&, \text{$i=n+1$ and $j=n+1$}
\end{cases}
\end{align*}
and the vector $b\in\mathbb{R}^{n+1}, b = [0, \dots, 0, -1]^T$.

By solving the system $B c = b$, the coefficients $c$ for extrapolating the Fock matrix derivatives can be obtained. 
Notice that, by construction, this solves the electronic and nuclear equations with the same set of coefficients.
The extrapolated Fock matrix derivatives are then constructed by linear combination of the previous matrix derivatives:
\begin{align}
    \frac{dF^{\text{extr.}}}{dR} = \sum_i c_i \frac{dF^{(i)}}{dR}
\end{align}

\vspace{1em}

\section{Computational complexity of the different gradient methods}
To asymptotically classify the different gradient methods, we assume that the (NEO-)Hartree-Fock calculation  is converged, so that that the MO coefficients and the MO transformed integrals $h^{(e)/(n)}_{pq}$, $S^{(e)/(n)}_{pq}$ and $g^{(e)/(n)}_{pqrs}$ are given. This analysis only takes into account the complexity for the gradient in one nuclear degree of freedom. The complexity for the calculation of all gradients is then scaled by $N$, the number of nuclear degrees of freedom.

\subsection{Analytical derivative}

\begin{algorithm}
\caption{Single index transformation on a two-body integral}
\label{alg:eri_transformation}
\begin{algorithmic}
\Require $C_{\mu p}$, $g_{\mu \nu \sigma \lambda}$
\State $g_{p \nu \sigma \lambda} \gets 0$
\For {$\mu \in 1\dots M$}
\For {$\nu \in 1\dots M$}
\For {$\sigma \in 1\dots M$}
\For {$\lambda \in 1\dots M$}
\For {$p \in 1\dots M$}
\State $g_{p \nu \sigma \lambda} \gets C_{\mu p} g_{\mu \nu \sigma \lambda}$
\EndFor
\EndFor
\EndFor
\EndFor
\EndFor
\end{algorithmic}
\end{algorithm}

The most expensive part of the analytical derivative  Eq.~(\ref{eq:ana_deriv}) is the transformation of the two-body integrals to the MO basis. This transformation can optimally be written as four $\mathcal{O}(M^5)$ steps:
\begin{align*}
&\frac{dg_{pqrs}}{dR_{Ik}} =\nonumber\\ &\sum_{\mu} C_{\mu p} \left(\sum_{\nu} C_{\nu q} \left(\sum_{\lambda} C_{\lambda r} \left(\sum_{\sigma} C_{\sigma s} \frac{dg_{\mu \nu \lambda \sigma}}{dR_{Ik}} \right)\right)\right),
\end{align*}
where each sum has $\mathcal{O}(M^5)$ steps, as can be seen in Alg.~\ref{alg:eri_transformation}.

\subsection{Numerical differentiation}
Depending on the order of convergence of the finite-difference stencil, at least one additional (NEO)-HF calculation is required, where the transformation of the two-body integrals to the MO basis is again the most expensive operation with $\mathcal{O}(M^5)$ steps.

\subsection{NEOCPHF}
In the NEOCPHF, the derivative of the Fock matrix requires the two-body integral skeleton derivatives, the calculation of which requires $\mathcal{O}(M^5)$ steps.

\subsection{Automatic differentiation}
Let $f:\mathbb{R}^n \rightarrow \mathbb{R}^m$ be a differentiable function with time complexity $\mathcal{O}(n^p m^q)$, where $p,q \in \mathbb{R}$. Forward mode automatic differentiation calculates the Jacobian of $f$ with a time complexity of $\mathcal{O}(1)\mathcal{O}(n^p m^q) n$ \cite{Griewank2008}. In this case, $f$ is the (NEO-)HF calculation with subsequent MO transformation of the electronic and nuclear integrals. Thus, the time complexity of $f$ is $\mathcal{O}(M^5)$ and $n=1$, where we treat one nuclear degree of freedom as input to the procedure. The time complexity of the forward mode automatic differentiation is then $\mathcal{O}(1)\mathcal{O}(M^5)=\mathcal{O}(M^5)$.

\vspace{1em}

\section{Amplitude-dependence of vibrational frequency}
\label{sec:amplitude_dependence}
\begin{figure}
    \centering
    \includegraphics[width=0.5\textwidth]{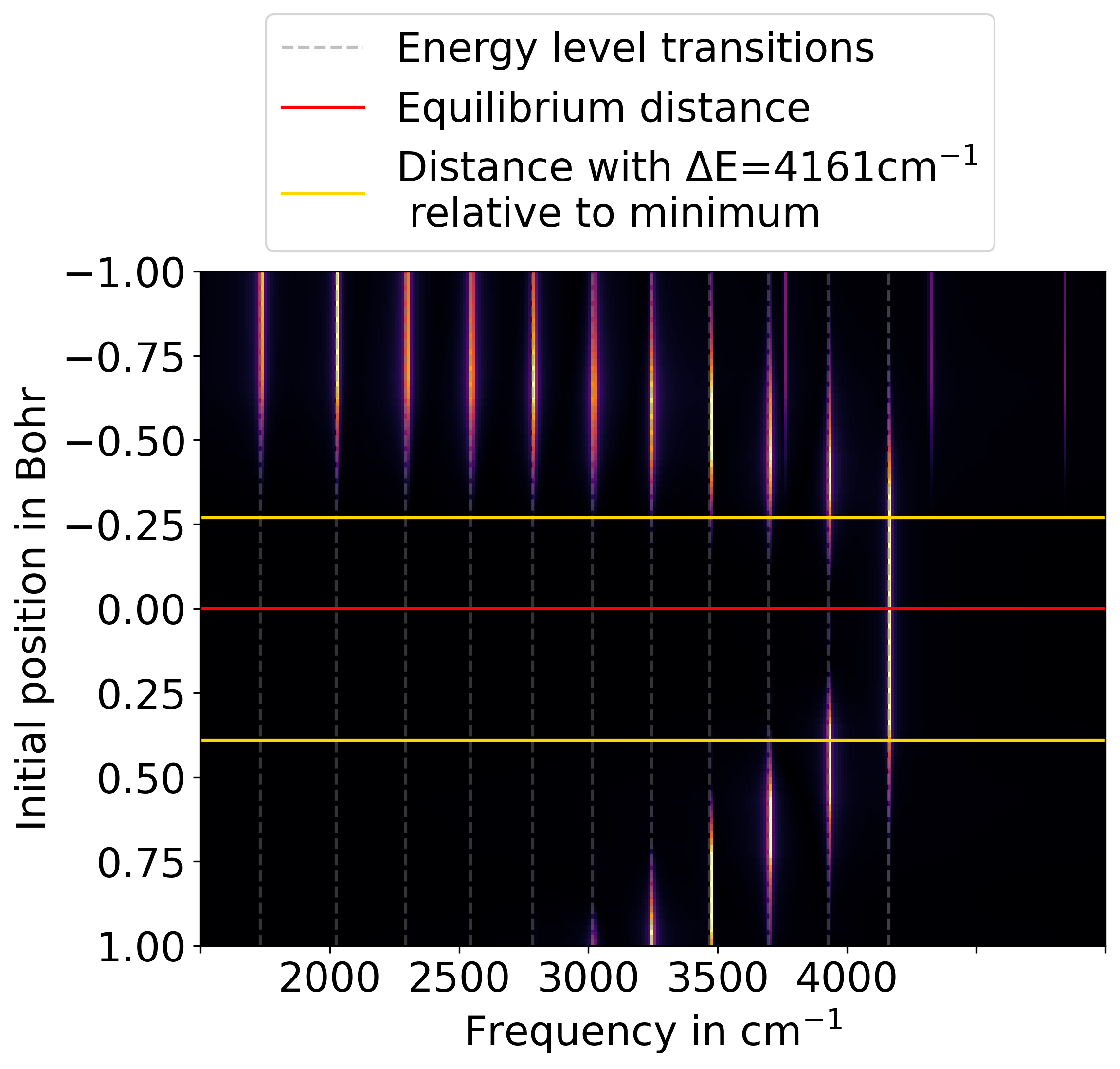}
    \caption{Spectral densities of a Gaussian wavepacket in the PES of a hydrogen molecule with varying initial displacements. The PES was calculated using FCI and the cc-pVQZ basis set. The red line indicates the equilibrium position and the yellow lines the displacements corresponding to an energy of \SI{4161}{\per \cm}.}
    \label{fig:h2_quantum_dynamics}
\end{figure}

Many molecules have anharmonic spectra, where the observed spectral lines are perturbed relative to the predictions obtained by using a harmonic oscillator model.

Our algorithm automatically takes into account anharmonicities as it considers ab-initio potential energy surfaces. 
An example for these simulated anharmonicities is given for \ce{H_2} in Fig.~\ref{fig:basisset_dependence_h2}.
This figure shows the estimated frequencies for the H--H stretching mode for different initial kinetic energies and for both the purely electronic and NEO methods.
For all simulation approaches, the predicted frequency decreases due to the anharmonicity in the PES of \ce{H_2}.

But it is not clear which initial energy and amplitude are the best choice for our MD simulation. 
While molecules are quantum objects, even the quantum nuclei in our algorithm are pinned to their classical expectation value.
It is therefore important to determine which initial amplitude in our simulation best matches the quantized motion of a molecule in low-energy excited states.

In order to answer this, we performed a classical quantum dynamics calculations, treating \ce{H_2} as a quantum oscillator in a potential energy landscape that was calculated using FCI and the cc-pVQZ basis set. 
We varied the initial position of a Gaussian wave packet and recorded the spectral densities of the resulting dynamics in Fig.~\ref{fig:h2_quantum_dynamics}.
These calculations show that a broad range of initial offsets lead to the fundamental frequency being dominant, while harmonics become dominant at larger amplitudes.

We conclude that there is a relatively free choice of initial displacement as long as it is sufficiently small.
From the \ce{H_2} calculation, an upper bound for the displacement seems to be around \SI{0.25}{Bohr}, which translates to an initial energy of \SI{5}{\milli \eV}.

\vspace{1em}

\section{Basis set dependence of NEO calculations}
\label{sec:app_basis_set_dependence_neo}
\begin{figure}
    \centering
    \includegraphics[width=0.5\textwidth]{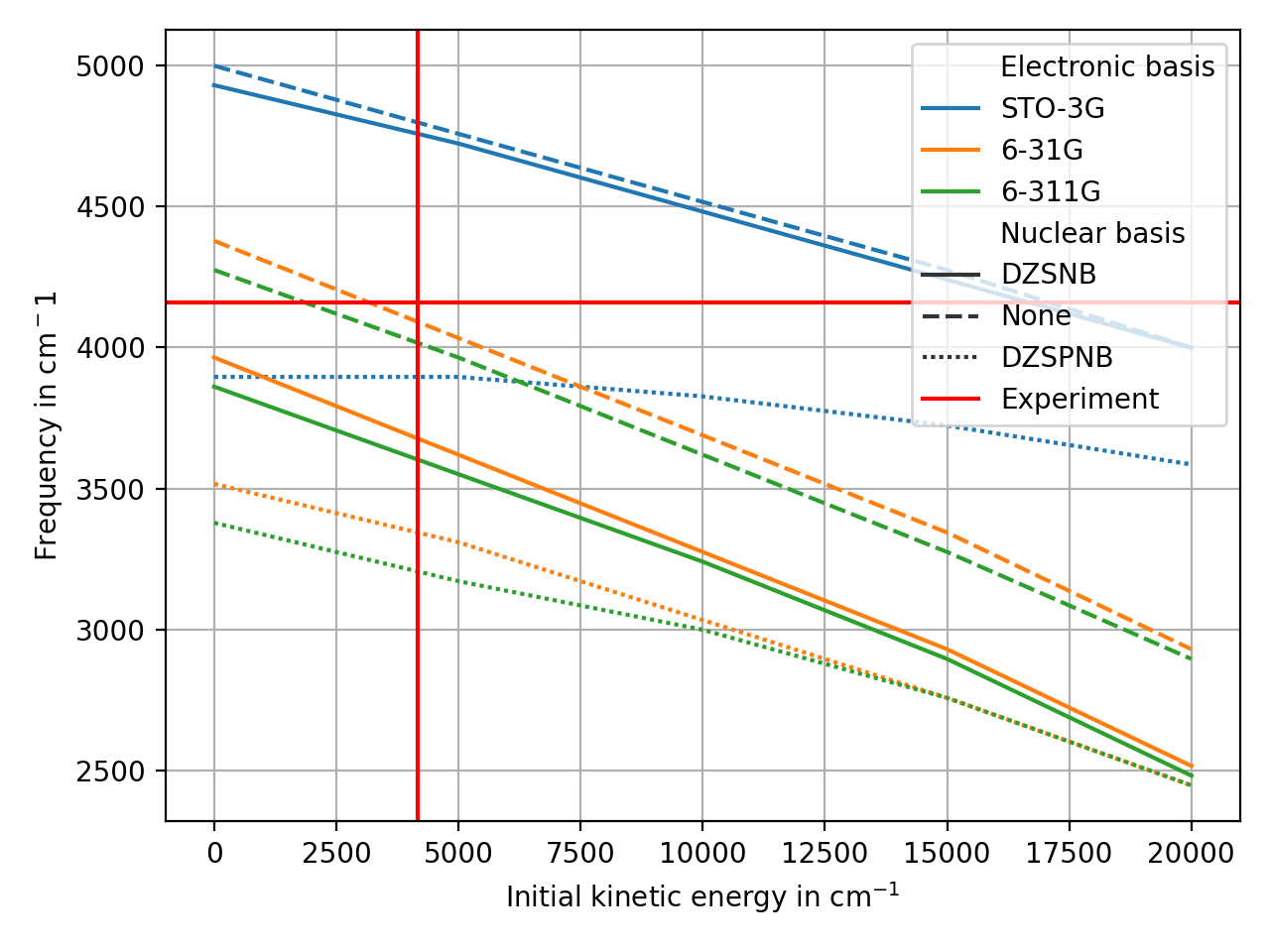}
    \caption{Calculated frequencies for performing MD with different initial displacements and basis sets. The red lines mark the experimental frequency of \SI{4161}{\per \cm}, while blue, orange and green mark the usage of the STO-3G, 6-31G and 6-311G electronic basis set and a dashed, solid and dotted line indicates the usage of no nuclear basis set, the DZSNB and the DZSPNB nuclear basis set, respectively.}
    \label{fig:basisset_dependence_h2}
\end{figure}

\begin{figure}
    \centering
    \includegraphics[width=0.5\textwidth]{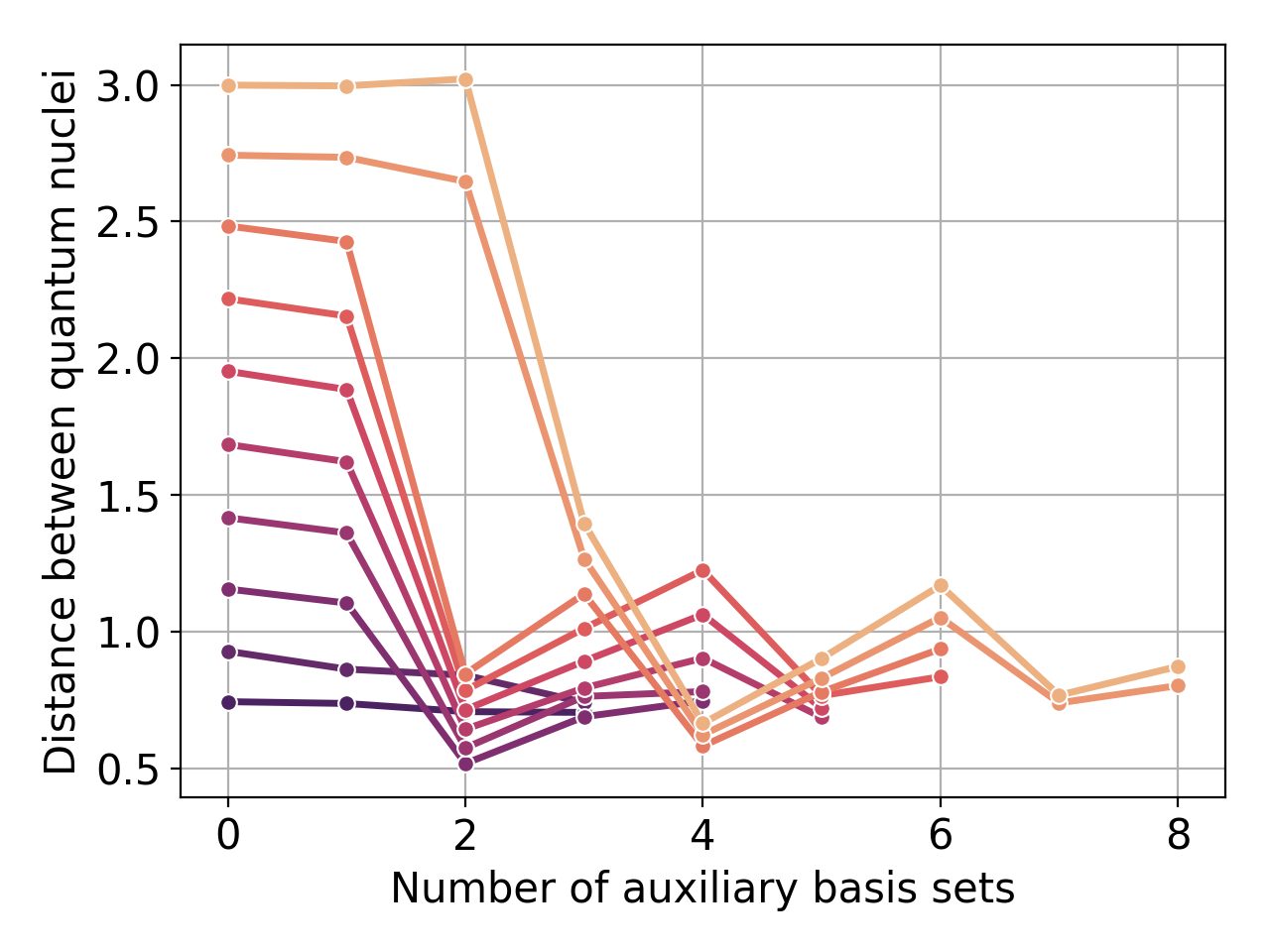}
    \caption{Effect of basis set size on internuclear distance in \ce{H_2}. The x-axis shows the number of nuclear auxiliary functions, which were equally spaced between the two initial basis set centers. The y-axis indicates the internuclear distance computed from the NEORHF nuclear density. The lines of different colors belong to different initial distances.}
    \label{fig:basisset_convergence_h2}
\end{figure}

We are working with relatively small basis sets, especially for the quantum nuclei.
This is because we noticed that with larger nuclear basis sets, the predicted frequencies severely underestimated the experimental frequencies.
An example of this is shown for \ce{H2} in Fig.~\ref{fig:basisset_dependence_h2}.

To elucidate this, we performed NEOSCF calculations on \ce{H_2}, placing an increasing number of auxiliary nuclear bases along the bond axis.
The expectation value of the distance between the two quantum nuclei is plotted in Fig.~\ref{fig:basisset_convergence_h2} as a function of the number of auxiliary nuclear basis functions. 
It can be seen that the internuclear distance quickly approaches the equilibrium distance, almost unaffected by the initial distance.
A similar, albeit less pronounced effect can be observed when using nuclear basis sets including higher angular momenta instead of using auxiliary basis functions.

\vspace{1em}

\section{Reduced NEOUCCSD}
\label{sec:reduced_neouccsd}

In order to shorten the NEOUCCSD ansatz for the Zundel simulation using the $p_z$-type nuclear orbitals, we performed VQEs using the full NEOUCCSD ansatz on 80 geometries sampled from a purely electronic Zundel MD simulation.
While these VQEs did not converge fully, we used them to obtain statistics for the 63 parameters of this NEOUCCSD ansatz, in particular the mean value and the standard deviation of the optimal values.
To judge the importance of each parameter, we used the sum of the absolute mean value and the standard deviation as a heuristic importance score.

In the UCCSD ansatz, each parameter can be identified with either a single or a double excitation. 
By incrementally adding the most important excitations to an initially empty UCCSD operator pool, we could construct increasingly refined ansatzes.
After adding a new excitation to the pool, we re-optimized the ansatz with a VQE and recorded the improvement of the recovered ground state energy.
After the 17 most important excitations were added, we noted a that further addition of excitation operators to the NEOUCCSD pool did not yield significant improvements to the energy.
Thus, the reduced NEOUCCSD ansatz contains these 17 excitations as a compromise between shallowness and expressivity of the quantum circuit.

\vspace{1em}

\section{Other Simulated Spectra}
\label{sec:app_more_simulated_spectra}
\begin{figure}
    \centering
    \includegraphics[width=\linewidth]{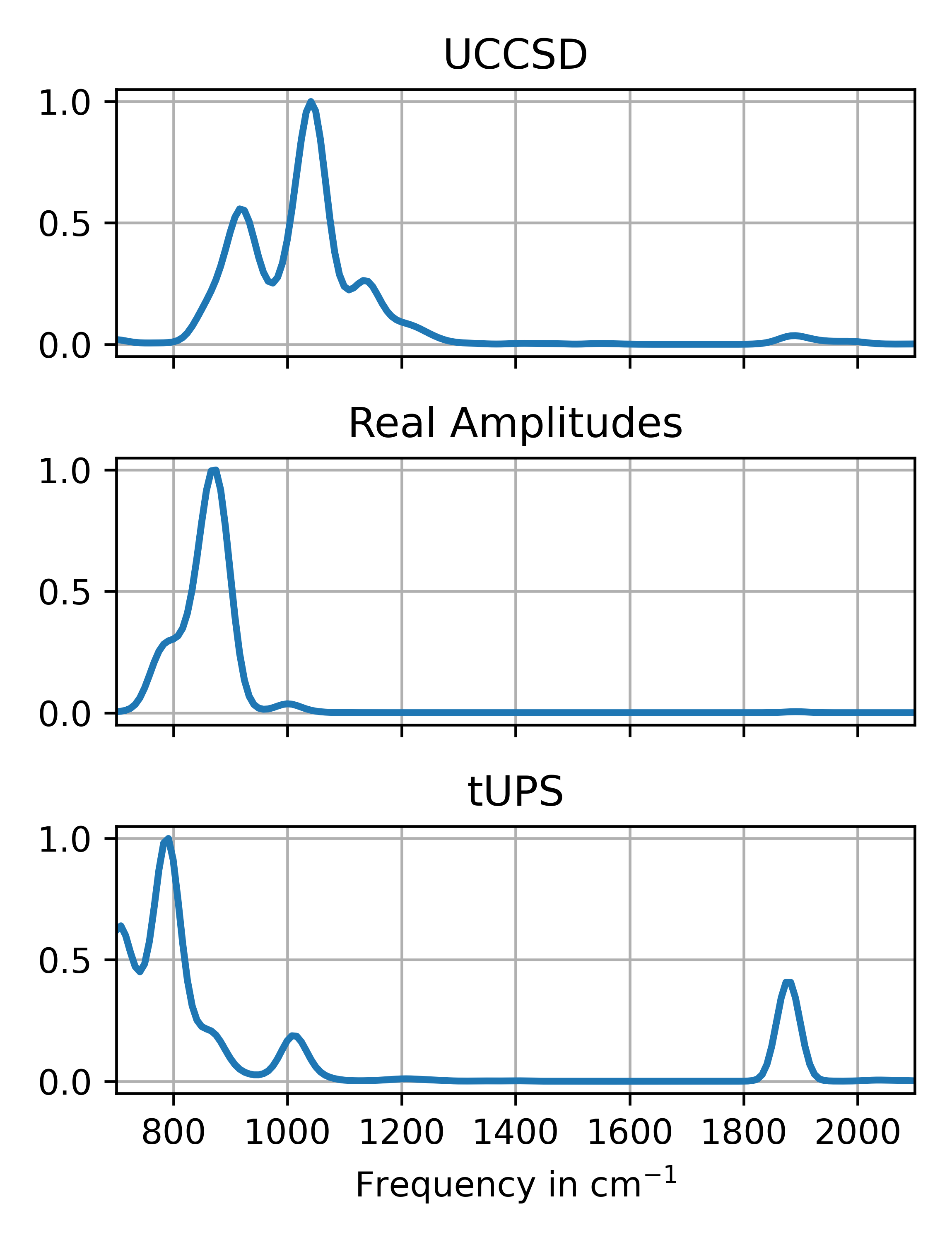}
    \caption{Computed spectra from purely electronic simulations of the Zundel proton transfer motion for the different ansatz types. The real amplitudes and tUPS ansatz had three layers each.}
    \label{fig:zundel_spectra_ele_hea}
\end{figure}

\begin{figure}
    \centering
    \includegraphics[width=\linewidth]{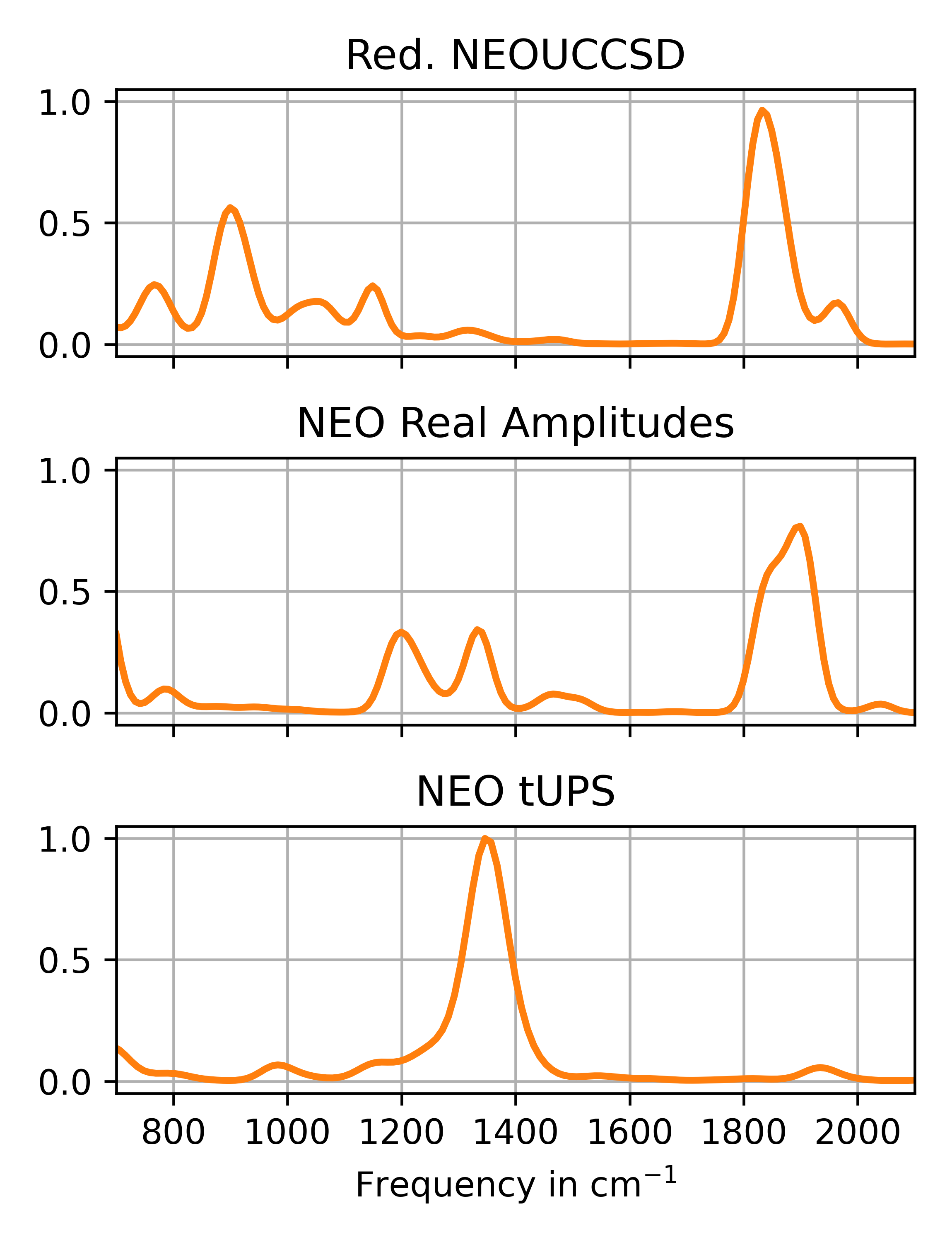}
    \caption{Computed spectra from NEO simulations for the Zundel proton transfer motion for the different ansatz types. The real amplitudes and tUPS ansatz had two layers each.}
    \label{fig:zundel_spectra_neo_hea}
\end{figure}

After determination of [5,8,14,23,25] as the best-performing active space for the Zundel ion, we repeated the simulation of the proton transfer mode for both the purely electronic and the NEO approach with different ansatzes. 
The purely electronic results are shown in Fig.~\ref{fig:zundel_spectra_ele_hea} and the NEO results are shown in Fig.~\ref{fig:zundel_spectra_neo_hea}. 
The NEO simulations were performed using the DZSPNB nuclear basis set.
In both the purely electronic and the NEO case, the hardware efficient simulations produce spectra which do not match the structure of the experimental spectra well.

\vspace{1em}

\section{Equilibrium geometries}
\label{sec:app_equilibrium_geometries}
Listings \ref{lst:geom_h2_ele}, \ref{lst:geom_h2_neo}, \ref{lst:geom_h2o_ele}, \ref{lst:geom_h2o_neo}, \ref{lst:geom_zundel_ele} and \ref{lst:geom_zundel_neo} show the  equilibrium geometries of the molecules which were then driven to oscillation in the molecular dynamics simulations. 
The geometries were obtained by optimization with the FIRE2 algorithm \cite{Gunol2020}, where convergence was defined as the maximal force acting on the atoms being less than \SI{5}{\milli \eV \per \bohr}.
Energy and force calculations had the same parameters as in the molecular dynamics simulations, that is, the electronic basis set was 6-31G, the nuclear basis set for NEO calculations was DZSNB, the electronic active space for water was [1,2,3,5,6] and the electronic active space for the Zundel ion was [5,8,14,23,25] unless otherwise specified.
The groundstate problems required for finding the energy and force at each optimization step were solved with classical (NEO)FCI and (NEO)CASCI.

\onecolumn
\begin{lstlisting}[float=false,caption={Purely electronic hydrogen geometry.}\label{lst:geom_h2_ele}, float=*]
2
Energy: -1.151697936814 Ha
H       0.000000000000000      0.000000000000000     -0.373098981459456
H       0.000000000000000      0.000000000000000      0.373098981459455
\end{lstlisting}

\begin{lstlisting}[caption={NEO hydrogen geometry.}\label{lst:geom_h2_neo}, float=*]
2
Energy: -1.067394995008 Ha
H       0.000000000000000      0.000000000000000     -0.393461935740797
H       0.000000000000000      0.000000000000000      0.393461935740797
\end{lstlisting}

\begin{lstlisting}[caption={Purely electronic water equilibrium geometry.}\label{lst:geom_h2o_ele}, float=*]
3
Energy: -75.995746605 Ha
O       0.000000000000000      0.072730000000000      0.072730000000000
H       0.000000000000000      1.020980000000000     -0.093660000000000
H       0.000000000000000     -0.093660000000000      1.020980000000000
\end{lstlisting}

\begin{lstlisting}[caption={NEO water equilibrium geometry.}\label{lst:geom_h2o_neo}, float=*]
3
Energy: -75.911615616 Ha
O       0.000000000000000      0.100658476667891      0.100658620064758
H       0.000000000000000      1.069674845669429     -0.076623611624971
H       0.000000000000000     -0.076623562074198      1.069674745239864
\end{lstlisting}

\begin{lstlisting}[caption= {{Purely electronic Zundel cation equilibrium geometry, calculated with active space {[5,8,14,23,25]}.}}\label{lst:geom_zundel_ele}, float=*]
7
Energy: -4145.449740977202 eV
O       0.000039823199578     -0.000035283645152      1.190875210981365
H       0.751374529005791     -0.287889849156283      1.704668709025223
H      -0.751274809173095      0.287780908521376      1.704721020976788
H       0.000101813344242     -0.000025273338760     -0.000458610290125
O       0.000355895438680     -0.000086613952785     -1.195764907682697
H      -0.288418766642247     -0.750727540467337     -1.710016527323335
H       0.289406039853359      0.750480974504334     -1.709968785515201
\end{lstlisting}

\begin{lstlisting}[caption= {{NEO Zundel cation equilibrium geometry, calculated with active space {[5,8,14,23,25]} and $s$-type nuclear orbitals.}}\label{lst:geom_zundel_neo}, float=*]
7
Energy: -4145.449740977202 eV
O       0.000039823199578     -0.000035283645152      1.190875210981365
H       0.751374529005791     -0.287889849156283      1.704668709025223
H      -0.751274809173095      0.287780908521376      1.704721020976788
H       0.000101813344242     -0.000025273338760     -0.000458610290125
O       0.000355895438680     -0.000086613952785     -1.195764907682697
H      -0.288418766642247     -0.750727540467337     -1.710016527323335
H       0.289406039853359      0.750480974504334     -1.709968785515201
\end{lstlisting}
\twocolumn

\end{document}